\magnification=\magstep1
\tolerance=500
\bigskip
\rightline{29 May, 2019}
\bigskip
\centerline{\bf An Elementary Canonical Classical and Quantum Dynamics}
\centerline{\bf for}
\centerline{\bf General Relativity}
\bigskip
\centerline{ L.P. Horwitz}
\bigskip
\centerline{ School of Physics, Tel Aviv University, Ramat Aviv
69978, Israel}
\centerline{ Department of Physics, Bar Ilan University, Ramat Gan
52900, Israel}
\centerline{ Department of Physics, Ariel University, Ariel 40700, Israel}
\bigskip
\noindent email: larry@tauex.tau.ac.il
\bigskip
\noindent{\it Abstract}
\bigskip
\par A consistent (off-shell) canonical classical and quantum dynamics in the framework of special relativity was formulated by Stueckelberg in 1941, and generalized to many- body theory by Horwitz and Piron in 1973 (SHP). In this paper, this theory is embedded into the framework of general relativity (GR), here denoted by SHPGR. The canonical Poisson brackets of the SHP theory remain valid (invariant under local coordinate transformations) on the manifold of GR, and provide the basis for formulating a canonical quantum theory. A scalar product is defined for constructing the Hilbert space and a Hermitian momentum operator defined. The Fourier transform is defined, connecting momentum and coordinate representations. The potential which may occur in the SHP theory emerges as a spacetime scalar mass distribution in GR, and electromagnetism corresponds to a gauge field on the quantum mechanical SHPGR Hilbert space in both the single particle and many-body theory. A diffeomorphism covariant form of Newton's law is found as an immediate consequence of the canonical formulation of SHPGR.   We compute the classical evolution of the off shell mass on the orbit of a particle and the force on a particle and its energy at the Schwarzschild horizon.  The propagator for evolution of the one body quantum state is studied and a scattering theory on the manifold is worked out. 
\bigskip
\noindent{\it Classification PACS} 03.30.+p, 03.65.-w, 04.20 Cr, 04.60 Ds, 04.90.+e
\bigskip
\noindent{\it Running Title: Embedding of SHP Theory into General Relativity}
\bigskip
\noindent{\it Introduction}
\bigskip

\par The relativistic canonical Hamiltonian dynamics of Stueckelberg, Horwitz and Piron (SHP)[1] with scalar potential and gauge field interactions for single and many body theory can, by local coordinate transformation, be embedded into the framework of general relativity (GR). Some of the results of this embedding are studied in this paper.
\par The theory was originally formulated for a single particle by Stueckelberg in 1941 [2][3][4]. Stueckelberg  envisaged the motion of a particle along a world line in spacetime that can curve and turn to flow backward in time, resulting in the phenomenon of pair annihilation in classical dynamics. The world line was then described by an invariant monotonic parameter $\tau$. 
The theory was generalized by Horwitz and Piron in 1973 [5] (see also [1][6][7])to be applicable to many body systems by assuming that the parameter $\tau$ is universal (as for Newtonian time [8][9]), enabling them to solve the two body problem classically, and later, a quantum solution was found by Arshanksy and Horwitz [10][11][12], both for bound states and scattering theory. 
 \par Performing a coordinate transformation to general coordinates, along with the corresponding transformation of the momenta (the cotangent space of the original Minkowski manifold), we obtain, in this paper,  the SHP theory in a curved space of general coordinates and momenta with a canonical Hamilton-Lagrange (symplectic) structure. We shall refer to this generalization as SHPGR. We study here the classical dynamics of a particle in the neighborhood of the Schwarzschild radius and obtain the force on a particle and its energy in this neighborhood.  
 \par The invariance of the Poisson bracket under local coordinate transformations provides a basis for the canonical quantization of the theory, for which the evolution under $\tau$ is determined by the covariant form of the Stueckelberg-Schr\"odinger equation (see also [13][14][15]). The one particle propagator is worked out, and a scattering theory developed (see also [16][17][18] for a discussion of scattering theory in the framework of general relativity where scattering theory is used to reach GR). 
 The formulation is also generalized here to a $U(1)$ Abelian gauge theory (electromagnetism on the manifold), but can be extended to the non-Abelian case. This provides a fundamental derivation of the framework assumed by Horwitz, Gershon and Schiffer [19][20] in their discussion\footnote{*} {A non-Abelian gauge was discussed there, and  then an Abelian limit was taken, leaving a term that could cancel caustic singularities.}  of the Bekenstein-Sanders fields [21] introduced into the TeVeS theory of Bekenstein and Milgrom [22][23][24], a geometrical way of obtaining the MOND theory introduced by Milgrom [25][26][27] to explain the rotation curves of galaxies. The potential term, entering into the structure of the scalar factor introduced by Bekenstein and Milgrom [22][23][24], as pointed out in [19], may provide a representation of ``dark energy'' as well as a phenomenological description of stars or galaxies in collision. 
 \par Birrell and Davies [28] have discussed fields on curved spacetime, and considerable progress has been made, as discussed by Poisson [29], in the formulation of Hamiltonian dynamics of such dynamical fields using Lagrangian functionals associated with the curvature of spacetime. The approach used in this paper is fundamentally different in that it studies a canonical dynamics (both Hamiltonian and Lagrangian) of {\it particles} on a curved spacetime.  
\par This method is applied also to the many body case, for which the SHP Hamiltonian is a sum of terms quadratic in four momentum with a many body potential term. Each particle is assumed {\it locally} to move in a flat Minkowski space, the tangent space of the general manifold of motions at that point; these local motions can then be mapped at each point $x^\mu$ by coordinate transformation into the curvilinear coordinates reflecting the curvature induced by the Einstein equations.
\par Throughout most of our development, we assume a $\tau$ independent
background gravitational field; the local coordinate transformations from the flat Minkowski space to the curved space are taken to be independent of $\tau$, consistently with an energy momentum tensor that is $\tau$ independent. In a more dynamical setting, when the energy momentum tensor depends on $\tau$, the spacetime evolves nontrivially;
the transformations from the local Minkowski coordinates to the curved space coordinates then depend on $\tau$. We discuss this situtation in an Appendix; many of the results for the $\tau$ independent case remain (such as the Poisson bracket relations), but some relations, such as the orbit equations, are modified.  
\par
\bigskip
\noindent{\it 1. Single particle in external potential}
\bigskip
\par We write the SHP Hamiltonian [1] as
$$K= {1\over 2M} \eta^{\mu\nu} \pi_\mu \pi_\nu + V(\xi) \eqno(1.1)$$
where $\eta^{\mu\nu}$ is the flat Minkowski metric $(-+++)$ and $\pi_\mu, \xi^\mu$ are the spacetime canonical momenta and coordinates in the local tangent space, following Einstein's use of the equivalence principle.
\par The existence of a potential term (which may be a Lorentz scalar), representing non-gravitational forces, implies that the ``free fall'' condition is replaced by a local dynamics carried along by the free falling system (an additional force acting on the particle within  the ``elevator'' according to the coordinates in the tangent space). 
\par The canonical equations are
$$ {\dot \xi}^\mu = {\partial K \over \partial \pi_\mu} \ \ \ \ \ \ {\dot \pi}_\mu =- {\partial K\over \partial \xi^\mu} = -{\partial V\over \partial \xi^\mu}, \eqno(1.2)$$
where the dot here indicates ${d\over d\tau}$, with $\tau$ the invariant universal ``world time''.
Since then
$$\eqalign{{\dot \xi}^\mu &= {1 \over M} \eta^{\mu \nu} \pi_\nu,\cr
 {\rm or} \ \ \ \pi_\nu &=\eta_{\nu \mu} M{\dot \xi}^\mu, \cr}
       \eqno(1.3)$$

the Hamiltonian can then be written as
$$K= {M\over 2} \eta_{\mu\nu}{\dot \xi}^\mu {\dot\xi}^\nu + V(\xi). \eqno(1.4)$$
\par It is important to note that, as clear from $(1.3)$, that ${\dot \xi}^0= {dt\over d\tau}$ has a sign {\it opposite} to $\pi_0$ which lies in the cotangent space of the manifold, as we shall see in the Poisson bracket relations below. The energy of the particle for a normal timelike particle should be positive (negative energy would correspond to an antiparticle [2][3][4][5][6][7]). The {\it physical momenta and energy} therefore correspond to the mapping
$$ \pi^\mu = \eta^{\mu\nu}\pi_\mu, \eqno (1.5)$$
back to the tangent space.  Thus, equivalently, from $(1.2)$, ${\dot \xi}^\mu =  (1/M) \pi^\mu$. This simple observation will be important in the discussion below of dynamics of a particle in the framework of general relativity, for which the metric tensor is non-trivial,  which we will discuss below.
\par We now transform the local coordinates (contravariantly) according to the diffeomorphism
$$ d\xi^\mu = {\partial \xi^\mu \over \partial x^\lambda} dx^\lambda \eqno(1.6)$$
to attach small changes in $\xi$ to corresponding small changes in the coordinates $x$ on the curved space, so that
$$ {\dot \xi}^\mu = {\partial \xi^\mu \over \partial x^\lambda} {\dot x}^\lambda . \eqno(1.7)$$
The Hamiltonian then becomes
$$K =  {M\over 2} g_{\mu\nu}{\dot x}^\mu {\dot x}^\nu + V(x), \eqno(1.8)$$
where $V(x)$ is the potential at the point $\xi$ corresponding to the point $x$ (actually the function $V(\xi)$ could be labelled $V_x(\xi)$, a function of $\xi$ in a small neighborhood of the point $x$), and
$$ g_{\mu\nu}= \eta_{\lambda \sigma}
 {\partial \xi^\lambda \over \partial x^\mu}{\partial \xi^\sigma \over \partial x^\nu} \eqno(1.9)$$ 
Since $V$ has significance as the source of a force in the local frame only through its derivatives, we can make this pointwise correspondence with a globally defined scalar function $V(x)$.\footnote{*}{Since $V(x)$ has dimension of mass, one can think of this function as a scalar mass field, reflecting forces acting in the local tangent space at each point.  It may play the role of ``dark energy'' [19][20]. If $V=0$, our discussion reduces to that of the usual general relativity, but with a well-defined canonical momentum variable.} We shall assume in most of the work of this paper that the geometric structure does not depend on $\tau$, and is concerned with the study of the covariant dynamical evolution of a system in a background gravitational field. We study the case of a $\tau$ dependent metric in the Appendix. 
\par The corresponding Lagrangian is then
$$L =  {M\over 2} g_{\mu\nu}{\dot x}^\mu {\dot x}^\nu - V(x), \eqno(1.10)$$
\par In the locally flat coordinates in the neighborhood of $x^\mu$, the symplectic structure of Hamiltonian mechanics ( {\it e.g.} da Silva [30] ) implies that the momentum \footnote{**}{We shall call the quantity $\pi_\mu$ in the cotangent space a {\it canonical momentum}, although it must be understood that its map back to the tangent space $\pi^\mu$ corresponds to the actual physically measureable momentum.}   $\pi_\mu$, lying in the cotangent space of the manifold $\{\xi^\mu \}$, transforms covariantly under the local transformation $(1.5)$, {\it i.e.}, as does ${\partial \over \partial \xi^\mu}$, so that we may define
$$ p_\mu = {\partial \xi^\lambda \over \partial x^\mu} \pi_\lambda. \eqno(1.11)$$
This definition is consistent with the transformation properties of the momentum defined by the Lagrangian $(1.10)$:
$$ p_\mu = {\partial L(x,{\dot x}) \over \partial {\dot x}^\mu}, \eqno(1.12)$$
yielding
$$ p_\mu = M  g_{\mu\nu}{\dot x}^\nu. \eqno(1.13)$$
\par We shall return to this point below, but continue here to study the effect of general coordinate transformations on the canonical variables $\xi^\mu$ and $\pi_\mu$.
\par The second factor in the definition $(1.9)$ of $g_{\mu\nu}$ in $(1.13)$ acts on ${\dot x}^\nu$; with $(1.7)$ we then have (as in $(1.11)$)
$$\eqalign{ p_\mu &= M  \eta_{\lambda \sigma} {\partial \xi^\lambda \over \partial x^\mu}{\dot \xi}^\sigma\cr
&= {\partial \xi^\lambda \over \partial x^\mu} \pi_\lambda . \cr} \eqno(1.14)$$
\par As we have remarked above for the locally flat space in $(1.5)$, the {\it physical} energy and momenta are given, according
to the mapping,
$$ p^\mu = g^{\mu\nu} p_\nu = M {\dot x}^\nu \eqno(1.15) $$
back to the tangent space of the manifold, which also follows directly from the local coordinate transformation of $(1.3)$ and $(1.5)$.
\par It is therefore evident from $(1.15)$ that
$$ {\dot p}^\mu = M {\ddot x}^\mu.  \eqno(1.16)$$
\par We see that ${\dot p}^\mu$, which should be interpreted as the force acting on the particle, is proportional to the {\it acceleration along the orbit of motion} (a covariant derivative plus a gradient of the potential), described by the geodesic type relation we discuss below.  This Newtonian type relation in the general manifold reduces in the limit of a flat Minkowski space to the corresponding SHP dynamics, and in the nonrelativistic limit, to the classical Newton law. We remark that this result does not require taking a post Newtonian limit, the usual method of obtaining Newton's law from GR.    
\par We now discuss the geodesic equation obtained by studying the condition
$${\ddot \xi}^\mu = - {1 \over M} {\dot \pi}_\mu = - {1 \over M}\eta^{\mu \nu} {\partial V(\xi) \over \partial \xi^\nu} . \eqno(1.17)$$
To do this, we compute \footnote{*}{Here we assume $\xi^\mu$ locally a function of $x(\tau)$ only.  If the spacetime evolves ($\tau$ dependent energy momentum tensor), then it is an explicit function of $\tau$ as well, {\it i.e.}, $\xi(x(\tau), \tau)$.  We show in the Appendix how ${\partial \xi(x(\tau),\tau) \over \partial \tau}$ is related to ${\partial g_{\mu\nu}(x(\tau),\tau) \over \partial \tau}$. We also recall here that, in the embedding, $V(x)$ is assumed to be a world scalar function [20][21].}   
$$ \eqalign{{\ddot \xi}^\mu &= {d \over d\tau} \bigl(  {\partial \xi^\mu \over \partial x^\lambda}{\dot x}^\lambda \bigr)\cr
&= {\partial^2 \xi^\mu \over \partial x^\lambda \partial x^\gamma} {\dot x}^\gamma {\dot x}^\lambda \cr
&+  {\partial \xi^\mu \over \partial x^\lambda} {\ddot x}^\lambda \cr
&= - { 1\over M} \eta^{\mu\nu}{\partial x^\lambda\over \partial \xi^\nu}{\partial V(x)\over \partial x^\lambda} \cr} \eqno(1.18)$$
for $V(x) = V(x(\tau))$,  
so that, after multiplying by ${\partial x^\sigma \over \partial \xi^\mu}$ and summing over $\mu$, we obtain
$$\eqalign{ {\ddot x}^\sigma &= -{\partial x^\sigma \over \partial \xi^\mu} {\partial^2 \xi^\mu \over \partial x^\lambda \partial x^\gamma}  {\dot x}^\gamma {\dot x}^\lambda\cr
&- { 1\over M} \eta^{\mu\nu}{\partial x^\lambda \over \partial \xi^\nu}{\partial x^\sigma \over \partial \xi^\mu} {\partial V(x)\over \partial x^\lambda}\cr} .\eqno(1.19)$$
Finally, with $(1.9)$ and the usual definition of the connection
$$ {\Gamma^\sigma}_{\lambda \gamma} = {\partial x^\sigma \over \partial \xi^\mu} {\partial^2 \xi^\mu \over \partial x^\lambda \partial x^\gamma} \eqno(1.20)$$
we obtain the modified geodesic type equation
$$  {\ddot x}^\sigma = - {\Gamma^\sigma}_{\lambda \gamma} {\dot x}^\gamma {\dot x}^\lambda
  - { 1\over M} g^{\sigma \lambda} {\partial V(x)\over\partial x^\lambda}, \eqno(1.21)$$
from which we see that the derivative of the potential $V(\xi)$ is mapped, under this coordinate transformation into a force resulting in a modification of the acceleration along the geodesic-like curves, {\it i.e.}, $(1.16)$ now reads
$$ {\dot p}^\mu =  M {\ddot x}^\nu=  -M {\Gamma^\sigma}_{\lambda \gamma} {\dot x}^\gamma {\dot x}^\lambda
  - g^{\sigma \lambda} {\partial V(x)\over\partial x^\lambda} \eqno(1.22)$$
\par The procedure that we have carried out here provides a canonical dynamical structure for the motions in the curvilinear coordinates. We now remark that the Poisson bracket remains valid for the coordinates $\{x,p\}$. In the local coordinates $\{ \xi, \pi\}$, the $\tau$ derivative of a function $F(\xi,\pi)$ is
$$ \eqalign{{dF(\xi,\pi) \over d\tau} &= {\partial F(\xi,\pi)\over \partial\xi^\mu}{\dot \xi}^\mu +{\partial F(\xi,\pi)\over \partial\pi_\nu} {\dot \pi}_\mu\cr
&= {\partial F(\xi,\pi)\over \partial\xi^\mu}{\partial K \over \partial \pi_\mu} -{\partial F(\xi,\pi)\over \partial\pi_\mu} {\partial K \over \partial \xi^\nu}\cr
&\equiv [F,K]_{PB} (\xi,\pi).\cr} \eqno(1.23) $$           
If we replace in  this formula
$$\eqalign{ {\partial \over \partial\xi^\mu} &={\partial x^\lambda \over \partial \xi^\mu}{\partial \over \partial x^\lambda} \cr
{\partial \over \partial\pi_\mu} &= {\partial \xi^\mu \over \partial x^\lambda}{\partial \over \partial p_\lambda}, \cr} \eqno(1.24)$$
we immediately (as assured by the invariance of the Poisson bracket under local coordinate transformations) obtain
$${dF(\xi,\pi) \over d\tau} = {\partial F \over \partial x^\mu}{\partial K \over \partial p_\mu} -{\partial F\over \partial p_\mu} {\partial K \over \partial x^\nu}\equiv [F,K]_{PB} (x,p) \eqno(1.25)$$
In this definition of Poisson bracket we have, as for the $\xi^\mu, \pi_\nu$ relation,
$$ [x^\mu, p_\nu]_{PB}(x,p) = {\delta ^\mu}_\nu. \eqno(1.26)$$
 The Poisson bracket of $x^\mu$ with the (physical energy-momentum) tangent space variable $p^\mu$ has then the tensor form
$$ [x^\mu, p^\nu]_{PB}(x,p) = g^{\mu \nu}. \eqno(1.27)$$
In the flat space limit, this relation reduces to the SHP bracket,
$$ [\xi^\mu, \pi^\nu]_{PB}(\xi,\pi) = \eta^{\mu \nu}. \eqno(1.28)$$
\par Continuing our analysis with $p_\mu$ (we drop the $(x,p)$ label on the Poisson bracket henceforth),
$$ [p_\mu, F(x)]_{PB} = - {\partial F \over \partial x^\mu}, \eqno(1.29)$$
so that $p_\mu$ acts infinitesimally as the {\it generator of translation} along the coordinate curves and  
$$[x^\mu, F(p)]_{PB} =  {\partial F(p) \over \partial p_\mu}, \eqno(1.30)$$
so that $x^\mu$ is the generator of translations in $p_\mu$.  In the classical case, if $F(p)$ is a general function of $p^\mu$, we can write at some point $x$,
$$[x^\mu, F(p)]_{PB} = g^{\mu\nu}(x) {\partial F(p) \over \partial p^\nu}, \eqno(1.31)$$
but in the quantized form, the factor $g^{\mu\nu}(x)$ cannot be factored out from polynomials, so, as for Dirac's quantization procedure [31][32][33], some care is required.  
\par This structure clearly provides a phase space which could serve as the basis for statistical mechanics (which we leave to a later publication), and lends itself to the construction of a canonical quantum theory on the curved spacetime, as we discuss below.
\par  We now turn to a discussion of the dynamics introduced into the curved space by the procedure outlined above. Although $p_\mu$ is not the physical energy momentum
as we have emphasized above, $p_\mu$ has a simpler Poisson  bracket relation with $x^\mu$, and this variable and its dynamical evolution will useful in further development of the theory. 
\par We start by developing the relation between ${\dot p}_\mu$ and the geodesic equations for $x^\mu$, and show that the result agrees with the direct Hamilonian calculation. Although the result has been discussed above, the alternative approach below has intrinsic geometrical interest. 
\par Recall from $(1.13)$ that
$$p_\mu = M g_{\mu \lambda} {\dot x}^\lambda, $$
so that
$${\dot p}_\mu = M \bigl( {\partial g_{\mu \lambda} \over \partial x^\gamma} {\dot x}^\gamma {\dot x}^\lambda
+ g_{\mu \sigma} {\ddot x}^\sigma \bigr). \eqno(1.32)$$
Since, by Eq.$(1.21)$,
$$  {\ddot x}^\sigma = - {\Gamma^\sigma}_{\lambda \gamma} {\dot x}^\gamma {\dot x}^\lambda
  - { 1\over M} g^{\sigma \lambda} {\partial V(x)\over\partial x^\lambda},$$
Eq. $(1.32))$ becomes
$$ {\dot p}_\mu =-{\partial V(x)\over\partial x^\mu}  + M\bigl( {\partial g_{\mu \lambda} \over \partial x^\gamma} {\dot x}^\gamma {\dot x}^\lambda - g_{\mu\sigma} {\Gamma^\sigma}_{\lambda \gamma} {\dot x}^\gamma {\dot x}^\lambda\bigr). \eqno(1.33)$$
We now use the relation
$$ {\Gamma^\sigma}_{\lambda \gamma} = {1 \over 2} g^{\sigma \eta}\bigl({\partial g_{\eta \lambda} \over \partial x^\gamma}+{\partial g_{\eta \gamma} \over \partial x^\lambda}- {\partial g_{\lambda \gamma} \over \partial x^\eta}\bigr) \eqno(1.34)$$
to obtain
$$ {\dot p}_\mu =-{\partial V(x)\over\partial x^\mu}  + M\bigl( {\partial g_{\mu \lambda} \over \partial x^\gamma} {\dot x}^\gamma {\dot x}^\lambda -{1\over 2}\bigl({\partial g_{\mu \lambda} \over \partial x^\gamma}+{\partial g_{\mu \gamma} \over \partial x^\lambda}- {\partial g_{\lambda \gamma} \over \partial x^\mu}\bigr){\dot x}^\gamma {\dot x}^\lambda \bigr). \eqno(1.35)$$
The first term in the brackets with coefficient $M$, symmetrized  under multiplication by 
 ${\dot x}^\gamma {\dot x}^\lambda$, cancels the first two terms of the contribution from the connection form with the result\footnote{*}{The 3D form of the second term plays an important role in the study of (3D) Hamiltonian stability analysis in the work of Horwitz, Ben Zion, Lewkowicz and Levitan [34], where it is called the ``reduced connection'' form.}
$${\dot p}_\mu =-{\partial V(x)\over\partial x^\mu} +{ M\over 2} {\partial g_{\lambda \gamma} \over \partial x^\mu} {\dot x}^\gamma {\dot x}^\lambda. \eqno(1.36)$$
\par We now calculate directly the $\tau$ derivative of the relation $p^\mu = g^{\mu\nu} p_\nu$. Using the identity
$$ g_{\sigma \gamma} {\partial g^{\mu\gamma} \over \partial x^\lambda} g_{\mu\beta} = - {\partial g_{\gamma\beta} \over \partial x^\lambda}, \eqno(1.37)$$
we find,  after a straightforward calculation,
$${\dot p}^\mu = -g^{\mu\nu} {\partial V(x)\over\partial x^\nu}- M {\Gamma^\mu}_{\gamma\lambda} {\dot x}^\gamma {\dot x}^\lambda,
= M{\ddot x}^\mu, \eqno(1.38)$$
a force directly associated with the acceleration along the orbit, reflecting both the
effect induced from the local potential function along with the acceleration induced by the geometry. Since the derivative of the metric diverges in the neighborhood of the black hole solution, Eqs. $(1.36)$ and $(1.38)$ provide a direct interpretation of the geometrical configuration as resulting in a very large {\it force} on the particle at the black hole horizon as would be seen in coordinates of this type.\footnote{**}{We emphasize here that the coodinates for description of the system are observed quantities in the observer's laboratory, as for $t$ and ${\bf x}$ in SR. Therefore the force is an effect seen by the observer, and would emerge, for example, in electromagnetic interaction as in $(7.6)$, and as pressure in statistical mechanics (which wll be discussed in a succeeding publication). Coordinate covariance corresponds to observations in different experimental arrangements, for which the underlying physical laws should emerge in an equivalent form.  We do not study here Kruskal type coordinates [35] which do not display a singularity at the horizon, but also leave this topic for a future publication.}   
 \par We may also write $(1.36)$ in terms of the full connection form by noting that with $(1.9)$,
 $${\partial g_{\lambda \gamma} \over \partial x^\mu}= \eta_{\alpha \beta}\bigl( {\partial^2\xi^\alpha \over \partial x^\lambda\partial x^\mu} {\partial \xi^\beta \over \partial x^\gamma} +{\partial \xi^\alpha \over \partial x^\lambda} {\partial^2\xi^\beta \over \partial x^\gamma\partial x^\mu}\bigr). \eqno(1.39)$$
 Multiplying by ${\dot x}^\gamma {\dot x}^\lambda$, the two terms combine to give a factor of two. We then return to the original definition of $\Gamma$ in $(1.20)$ in the form
 $$ {\partial^2\xi^\alpha \over \partial x^\lambda\partial x^\mu}= {\partial \xi^\alpha \over \partial x^\sigma} {\Gamma^\sigma}_{\lambda \mu}, \eqno(1.40)$$
 so we can write 
 $$\eqalign{{\partial g_{\lambda \gamma} \over \partial x^\mu}{\dot x}^\gamma {\dot x}^\lambda
 &= 2\eta_{\alpha \beta} {\partial \xi^\alpha \over \partial x^\sigma} {\partial \xi^\beta \over \partial x^\gamma}
 {\Gamma^\sigma}_{\lambda \mu} {\dot x}^\gamma {\dot x}^\lambda \cr
 &= 2 g_{\sigma \gamma}{\Gamma^\sigma}_{\lambda \mu} {\dot x}^\gamma {\dot x}^\lambda.\cr} \eqno(1.41)$$
 We  therefore have
$${\dot p}_\mu =-{\partial V(x)\over\partial x^\mu}+M g_{\sigma\gamma}{\Gamma^\sigma}_{\lambda \mu} {\dot x}^\gamma {\dot x}^\lambda.
 \eqno(1.42)$$
\par We now return to the Hamiltonian $(1.8)$ and carry out the calculation directly. Since ${\dot x}^\mu$ is, in general,  a function of $x^\mu$, we write the Hamiltonian (using $(1.13)$) in terms of the momenta, assured by the canonical structure to be independent variables,
$$K = {1 \over 2M} g^{\alpha \beta} p_\alpha p_\beta  + V(x). \eqno(1.43)$$
Then,
$${\dot p}_\mu = - {1 \over 2M} {\partial g^{\alpha \beta} \over \partial x^\mu}p_\alpha p_\beta - {\partial V(x)\over \partial x^\mu}. \eqno(1.44)$$
Returning to the form with ${\dot x}$ again, we have, with $p_\alpha = M g_{\alpha \lambda} {\dot x}^\lambda$,
$${\dot p}_\mu = {M\over 2} {\partial g^{\alpha \beta} \over \partial x^\mu} g_{\alpha \lambda}g_{\beta \gamma}{\dot x}^\gamma {\dot x}^\lambda  -{\partial V(x)\over \partial x^\mu}. \eqno(1.45)$$
Using, however, the  identity $(1.38)$, we recover the form obtained in $(1.37)$. In Section 3, we apply these results to study the dynamics of a particle near the Schwarzschild horizon [36], but first discuss the stability of the off-shell mass in the following. 
\bigskip
\noindent{\it 2. Off Shell Mass Evolution}
\bigskip
\par In this section, we consider the variation of the measured mass of a particle as it moves along along its orbit in the manifold.
\par In Eq. $(1.1)$, the potential term $V(\xi)$ generates a force through the Hamilton equations. The mass of the particle that is actually  measured in the laboratory is defined by
$$ m^2 =- \eta^{\mu\nu} \pi_\mu \pi_\nu \eqno(2.1)$$
Its $\tau$ derivative is
$$ \eqalign{{dm^2 \over d\tau} &= -2\eta^{\mu\nu} \pi_\mu {\dot \pi}_\nu\cr
 &= +M {\dot \xi}^\nu{\partial V(\xi) \over \partial \xi}. \cr} \eqno(2.2)$$
 The derivative of the potential function therefore induces a change in the particle mass due to interaction. This change in the effective mass of a particle in interaction could, for example, account phenomenologically for the transition to stability of the neutron in the nucleus. This was one of the motivations for constructing the SHP theory [1][5][6][7]).
 \par It follows directly from $(2.2)$ and the local coordinate transformation that the same result is true on the manifold of SHPGR. To see how this follows within the framework of SHPGR, we start with the transformation of $(2.1)$ to the curved spacetime. With $(1.13)$ we obtain
$$ m^2 = -g^{\lambda \sigma} p_\lambda p_\sigma \equiv -g_{\lambda\sigma}p^\lambda p^\sigma. \eqno(2.3)$$
From $(1.36)$ and $(1.13)$ we see that
$$ \eqalign{ {dm^2 \over d\tau} &=- {\partial g^{\lambda\sigma} \over \partial x^\gamma} {\dot x}^\gamma p_\lambda p_\sigma -2 g^{\lambda \sigma} {\dot p}_\lambda p_\sigma \cr
&= - {\partial g^{\lambda\sigma} \over \partial x^\gamma}{ 1 \over M}g^{\gamma \mu} p_\mu  p_\lambda p_\sigma 
 - 2 g^{\lambda\sigma} p_\sigma [-{\partial V \over \partial x^\lambda} + {M \over 2}{\partial g_{\sigma \kappa} \over \partial x^\lambda} {1\over M} g^{\sigma \alpha} p_\alpha g^{\kappa \beta} p_\beta ]. \cr} \eqno(2.4)$$
Finally, again using the formula
$${\partial g_{\sigma \kappa} \over \partial x^\lambda} g^{\sigma \alpha} g^{\kappa\beta} = -{\partial g^{\alpha\beta} \over \partial x^\lambda}, \eqno(2.5)$$
we see that the first and last terms cancel and we are left with the SHPGR form of $(2.2)$, i.e.,
$$ {dm^2 \over d\tau} = 2 g^{\lambda \sigma} p_\sigma{\partial V \over \partial x^\lambda} \eqno(2.6)$$
or, in terms of velocity along the orbit,
$$ {dm^2 \over d\tau} = 2 M {\dot x}^\lambda{\partial V \over \partial x^\lambda}, \eqno(2.7)$$
so that
$${dm \over d\tau} = { M\over m} {\dot x}^\lambda{\partial V \over \partial x^\lambda} . \eqno(2.8)$$
For far off shell particles, for $m$ small, so that $M/m$ is large, a small potential gradient can have a large effect on the mass variation (recall that this potential may represent the ``dark energy'' distribution, and may occur implicitly in the MOND formulas[19][20]). In the absence of a non-constant potential term, the off-shell mass would be striclty conserved along the orbit.
\bigskip
\noindent {\it 3. Dynamics of a Particle Near the Schwarzschild Horizon}
\bigskip
\par It is well-known from the geodesic equation, in standard GR, that a particle near the Schwarzschild radius undergoes a very large acceleration, which can be thought of as due to a large force. With the formula $(1.38)$ we can compute directly the force, as the $\tau$ derivative of the momentum in the radial direction and, by computing ${\dot x}^0$, the redshift. We shall also compute the particle energy ($E= p^0$) in this region.
In this study, we take $V=0$; it may in this sense be considered as a property of the SHPGR phase space in GR. The equations of motion in this case take on the usual geodesic form.\footnote{*}{The large geodesic acceleration of GR is, by $(1.38)$, interpreted here directly in terms of a force.} 
\par We consider the symmetric case of the Schwarzschild solution [36] for which only the $r,t$ ($\mu = 1,0$) components are relevant. Let us call $ {\dot x}^\mu = v^\mu$, so that the geodesic equation 
for the $0$ component becomes
$$ \eqalign{{dv^0 \over d\tau} &= -{\Gamma^0}_{\lambda \sigma} v^\lambda v^\sigma\cr
&= - {1 \over 2} g^{00}\bigl\{ {\partial g_{\sigma 0}\over \partial x^\lambda} +
{\partial g_{\lambda 0}\over \partial x^\sigma} - {\partial g_{\sigma \lambda}\over \partial x^0}\bigr\}v^\lambda v^\sigma. \cr} \eqno(3.1)$$
Taking into account that for the (diagonal) Schwarzschild metric
$$ g_{00} = - \bigl( 1 -{2M_S G\over r}\bigr) = - {1 \over g_{11}}, \eqno(3.2)$$
where $M_S$ is the black hole mass, we obtain
$${dv^0 \over d\tau}= - g_{00} {\partial g_{00} \over \partial x^1} v^1 v^0. \eqno(3.3)$$
Since ${\partial g_{00} \over \partial x^1} v^1= {dg_{00}\over d\tau}$, we can write this result as
  $${dv^0 \over d\tau} = - g^{00}{dg_{00} \over d\tau} v^0 $$
or, since  ( $g^{00} = {1\over g_{00}})$
$${d \over d\tau} (g_{00} v^0) =0$$
we have that (Dirac [37])
$$ g_{00} v^0 = k ={\rm constant \ \ in} \ \tau. \eqno(3.4)$$
Since for the Schwarzschild metric [37], the Hamilonian has the form
$$K= {1\over 2M} (g^{00} {p_0}^2 + g^{11} {p_1}^2), \eqno(3.5)$$
one obtains
$$ {dt \over d\tau} = v^0 = {\partial K \over \partial p_0} = {1 \over M} g^{00} p_0 \eqno(3.6) $$
or
$$  g_{00} v^0 = {p_0 \over M}= k. \eqno(3.7)$$
Since $g_{00}<0$, and $v^0 >0$, it follows that $k<0$.
From $(3.6)$, we see that then

$$v^0 = {dt\over d\tau} =- \bigl( 1 -{2M_s G\over r}\bigr)^{-1} k >0. \eqno(3.8)$$
\par Although  the kinematic phase space ($p_\mu$ is in the cotangent bundle of the manifold) is the set $\{ x^\mu, p_\mu\}$, we recognize, as pointed out above, that it is $p^0$ that has the interpretation of the energy of the particle and $p^i, \ \ i=1,2,3$ have the interpretation of physical momenta (note from $(3.7)$ that $p_0$ is a constant of the motion and is negative).
\par The energy of the particle is then
$$E= p^0= g^{00}p_0  = g^{00} M k>0. \eqno(3.9)$$
\par It also follows from $(3.8)$ that in a finite increment of $\tau$, the corresponding increment of $t$ at the horizon undergoes an infinite redshift (as is generally obtained from the structure of the metric)\footnote{*}{ The usual argument leads to an infinite redshift relative to the proper time interval, defined by $ds^2 = -g_{\mu\nu} dx^\mu dx^\nu$. This quantity is, however, dynamical, as discussed in [1], and does not necessarily reflect the invariant evolution of the system as recorded in terms of an ideal universal laboratory clock ($\tau$). The redshift we obtain here (explicitly in $(3.16)$) is computed in terms of this absolute time, a universal time measured (ideally) in the rest frame of any inertial laboratory.}. 
\par Moreover, at $r \rightarrow \infty$,we see that $ k =- {v^0}_\infty $.
It follows from the Hamilton equations that, since $g_{00}\rightarrow -1$ at $r\rightarrow \infty$, 
$$ {v^0}_\infty= {\partial K \over \partial p_0}|_\infty=  {g^{00} p^0\over M}|_\infty  ={E_\infty \over M} \eqno(3.10)$$
\par Therefore,
$$ E =  \bigl( {1 \over 1-{2M_s G \over r} }\bigr) E_\infty \eqno(3.11)$$
Solving for $r$ for a given $E$, we have
$$ r(E) = {2M_sG \over 1 + {E_\infty \over E}}. \eqno(3.12)$$
\par Although $ p_0$ (the generator of translations in $t=x^0$) is constant in $\tau$, the energy of the particle $E= p^0$ grows rapidly towards the surface of the black hole.
Particle production at high energies ($E>>E_\infty$) could therefore be induced close to the horizon, as assumed by Hawking[38].
\par We now turn to calculate the force on the particle close to the horizon.
It is of interest (and useful) to first calculate the rate of change of the cotangent space variable ${\dot p}_1$, more singular than the physical ${\dot p}^1$.
\par From the general relation $(1.36)$, in the radial direction, for $V=0$ or constant,
$$ \eqalign{{\dot p}_1 &= {M \over 2} {\partial g_{\lambda\gamma} \over \partial x^1} {\dot x}^\gamma{\dot x}^\lambda \cr
&= {M \over 2} \{ {\partial g_{00} \over \partial x^1} ({\dot x}^0)^2 +
{\partial g_{11} \over \partial x^1} ({\dot x}^1)^2 \}. \cr} \eqno (3.13)$$
We first evaluate $(v_1)^2$. From $(2.3)$, in the Schwarzschild metric, we have 
$$ {m^2 \over M^2} = {1 \over g_{00}} ( (v^1)^2 - {g_{00}}^2 (v^0)^2), \eqno(3.14)$$
so that
$$ (v^1)^2 = g_{00} {m^2 \over M^2} + (g_{00} v^0)^2. \eqno(3.15)$$
\par From $(3.8)$ ( ${\dot x}^0 \equiv v^0$) and the result $k =- {v^0}_\infty$, we obtain
$$ v^0 = {{v^0}_\infty\over ( 1 -{2M_s G\over r})}. \eqno(3.16)$$
Then using the explicit forms $(3.2)$ for the metric, we obtain 
$${\dot p}_1 = - {MM_s G \over r^2} \bigl\{ {2({v^0}_\infty)^2 \over \bigl( 1 -{2M_s G\over r}\bigr)^2} - {1 \over \bigl( 1 -{2M_s G\over r}\bigr)}{m^2\over M^2} \bigr\}.\eqno(3.17)$$
\par For $r= 2M_sG + \epsilon$, for $\epsilon <<2M_sG$,
$$ 1- {2M_sG \over r}\cong {\epsilon \over 2 M_sG}, \eqno(3.18)$$ 
and finite ${m^2 \over M^2}$,
$${\dot p}_1 \cong -2 {MM_s G \over \epsilon^2} ({v^0}_\infty)^2. \eqno(3.19)$$
\par Since classically ${\dot p}_1$ (and therefore $p_1$) grows rapidly, one might expect that the quantum mechanical dispersion in $p_1$ also becomes large, and therefore that (by the uncertainty relation) the particle becomes highly localized in the neighborhood of the horizon (see also [39]).
\par On the other hand, for $r \rightarrow \infty$, it follows from $(3.17)$ that
$${\dot p}_1 \cong - {MM_s G \over r^2} \{ 2({v^0}_\infty)^2 -{m^2 \over M^2}\}. \eqno(3.20)$$
Since $({v^0}_\infty)= {E_\infty\over M}$, $(3.20$ can be written as
$${\dot p}_1 \cong - {MM_s G \over r^2} \{ ({v^0}_\infty)^2 +{{E_\infty}^2-m^2 \over M^2}\}. \eqno(3.21)$$
For the free motion at $\infty$, we have by definition,$ {E_\infty}^2-m^2=
({\bf p}_\infty)^2 \geq 0$, and small, and $({v^0}_\infty)^2 \cong 1$, we see that $ {\dot p}_1$ is close to the Newtonian force. 
\par We now study the behavior of the physically observable  momentum at the horizon.  Since $p^1 = g^{11}  p_1$, we can relate ${\dot p}^1$ to our previous result for
${\dot p}_1$,
$${\dot p}^1 = {dg^{11}\over d\tau} p_1 + g^{11}{\dot p}_1. \eqno(3.22)$$
For $g^{11} = (g_{11})^{-1}$, and ${\dot r}\equiv {\dot x}^1\equiv v^1$, we find directly that\footnote{*}{This calculation is, of course, equivalent to using $(1.16)$, but is of interest in itself, showing the relation between ${\dot p}_1$ and ${\dot p}^1$.}
$${\dot p}^1 = - {2M_sG \over r^2} {1 \over \bigl( 1 -{2M_s G\over r}\bigr)^2} v^1 p_1 +
{1 \over \bigl( 1 -{2M_s G\over r}\bigr)} {\dot p}_1. \eqno(3.23)$$
Since $v^1 p^1 = M (v^1)^2$, we see that
$$ {\dot p}^1 = - {2M_sG \over r^2} {1 \over \bigl( 1 -{2M_s G\over r}\bigr)^3} (v^1)^2 + {1 \over \bigl( 1 -{2M_s G\over r}\bigr)} {\dot p}_1 \eqno(3.24)$$
Substituting the result for ${\dot p}_1$ and using $(3.15)$ for $(v^1)^2$, we obtain
$${\dot p}^1 = -{4MM_s G \over r^2} {1 \over \bigl( 1 -{2M_s G\over r}\bigr)^3}({v^0}_\infty)^2 + {3M M_s G \over r^2}{1 \over \bigl( 1 -{2M_s G\over r}\bigr)^2} {m^2 \over M^2} \eqno(3.25)$$
For $r\rightarrow \infty$, one finds
$$ {\dot p}^1= -{4MM_s G \over r^2}({v^0}_\infty)^2 + {3MM_s G \over r^2}{m^2 \over M^2}.{\dot p}^1 \eqno(3.26)$$
Furthermore, for $ ({v^0}_\infty)^2 = ({E_\infty\over M})^2= {{{\bf p}_\infty}^2 +m^2 
\over M^2} $, we obtain
$$ {\dot p}^1 = -{MM_s G \over r^2}\bigl({m^2 \over M^2}\bigr) -{4MM_s G \over r^2}{{{\bf p}_\infty}^2\over M^2}. \eqno(3.27)$$
As for ${\dot p}_1$, this result contains small differences from the standard Newtonian force.
\par For $r \rightarrow 2M_s G +\epsilon$, the general result $(3.25)$ yields
$$ {\dot p}^1 \approx -{8M(M_sG)^2 \over \epsilon^3} ({v^0}_\infty)^2 + 3 {m^2 M_sG \over M\epsilon^2}, \eqno(3.28)$$
a stronger divergence at the horizon than in ${\dot p}_1$. \footnote{*}{D. Momeni [40] has recently applied this theory to obtain exact solutions for the covariant classical and quantum oscillator (the quantized theory is treated here in Section 5) in the neighborhood of the black hole horizon.}  

  \par We now turn to  a discussion of the many body problem.

\bigskip
\noindent{\it 4. The many body system with interaction potential}
\bigskip
\par The many body Hamiltonian of the SHP theory is
$$K= \Sigma_{i=1}^N{1\over 2M_i} \eta^{\mu\nu} \pi_{\mu i} \pi_{\nu i} + V(\xi_1, \xi_2,\dots \xi_N),\eqno(4.1)$$
where the potential $V(\xi_1, \xi_2,\dots \xi_N)$ is a function of the locally flat coordinates in the neighborhood of each of the particles at $\{x_i\}$. Although this function is Lorentz scalar, Poincar\'e invariance is, in general, inapplicable (even in the two-body case), unless all of the particles are in a sufficiently small neighborhood to be able to neglect the effects of curvature.
\par The Hamilton equations are (in the tangent space in the neighborhood of each particle at the point $x_i$)
$$ {{\dot \xi}^\mu}_i = {\partial K \over \partial \pi_{\mu i}} \ \ \ \ \ \ {\dot \pi}_{\mu i} =- {\partial K\over \partial {\xi^\mu}_i} = -{\partial V\over \partial {\xi^\mu}_i}, \eqno(4.2)$$
We then have
$$\eqalign{{{\dot \xi}^\mu}_i &= {1 \over M_i} \eta^{\mu \nu} \pi_{\nu i},\cr
 {\rm or} \ \ \ \pi_{\nu i}  &=\eta_{\nu \mu} M_i{{\dot \xi}^\mu}_i, \cr} 
       \eqno(4.3)$$
\par Following the procedure we used for the one-body case above, we may substitute this expression into the Hamiltonian to obtain
$$K = \Sigma_{i=1}^N{M_i\over 2} \eta_{\mu \nu} {{\dot \xi}^\mu}_ i {{\dot \xi}^\nu}_i + V(\xi_1, \xi_2,\dots \xi_N),\eqno(4.4)$$
At the location ${x^\mu}_i $ of the $i^{th}$ particle, since in this neighborhood, $\xi^{\mu i}$ is a function locally of  ${x^\mu}_i$,  we can then make a local coordinate transformation
$$ d{\xi^\sigma}_i = {\partial {\xi^\sigma}_i \over \partial {x^\mu}_i } d{x^\mu}_i. \eqno(4.5)$$ 
Defining
$$ g_{\mu \nu} (x_i) =  \eta_{\sigma\lambda}{\partial {\xi^\sigma}_i \over \partial {x^\mu}_i }{\partial {\xi^\lambda}_i \over \partial {x^\nu}_i }, \eqno(4.6)$$
one obtains the Hamiltonian in terms of the four-velocities; changing notation for the arguments of the potential,\footnote{*}{We assume that (general covariance) $V(x_1, x_2,\dots x_N)$ is a scalar function under local diffeomorphisms of any of the variables.}      
$$K= \Sigma_{i=1}^N {M_i\over 2} g_{\mu \nu}(x_i) {{\dot x}^\mu}_i{{\dot x}^\nu}_i + V(x_1, x_2,\dots x_N),\eqno(4.7)$$
with corresponding Lagrangian
$$L= \Sigma_{i=1}^N{M_i\over 2} g_{\mu \nu}(x_i) {{\dot x}^\mu}_i{{\dot x}^\nu}_i - V(x_1, x_2,\dots x_N).\eqno(4.8)$$
\par As for the one body case, we can find the equations for the geodesic motion of the particles as follows. Since $(1.6)$ is valid for each of the particle coordinates, 
$$ {{\dot \xi}^\mu}_i = {\partial {\xi^\mu}_i \over \partial {x^\lambda}}_i {{\dot x}^\lambda}_i , \eqno(4.9)$$
from which we similarly obtain
$$ \eqalign{{{\ddot \xi}^\mu}_i &= {d \over d\tau} \bigl(  {\partial {\xi^\mu}_i \over \partial {x^\lambda}_i}{{\dot x}^\lambda}_i \bigr)\cr
&= {\partial^2 {\xi^\mu}_i \over \partial {x^\lambda}_i \partial {x^\gamma}_i} {{\dot x}^\gamma}_i {{\dot x}^\lambda}_i \cr
&+  {\partial {\xi^\mu}_i \over \partial {x^\lambda}_i} {{\ddot x}^\lambda}_i \cr
&= - { 1\over M_i} \eta^{\mu\nu}{\partial {x^\lambda}_i\over \partial {\xi^\nu}_i}{\partial V(x_1, x_2,\dots x_N)\over \partial {x^\lambda}_i}. \cr} \eqno(4.10)$$
\par The many body geodesic curves are therefore described by
$$\eqalign{ {{\ddot x}^\sigma}_i &= -{\partial {x^\sigma}_i \over \partial {\xi^\mu}_i} {\partial^2 {\xi^\mu}_i \over \partial {x^\lambda}_i \partial {x^\gamma}_i}  {{\dot x}^\gamma}_i {{\dot x}^\lambda}_i\cr
&- { 1\over M_i} \eta^{\mu\nu}{\partial {x^\lambda}_i \over \partial {\xi^\nu}_i}{\partial {x^\sigma}_i \over \partial {\xi^\mu}_i} {\partial V(x_1, x_2,\dots x_N)\over \partial {x^\lambda}_i } \cr} .\eqno(4.11)$$
We can consider the Jacobian for the local mapping $(4.5)$ as a {\it field}, a mapping defined over all $\{x^\mu \}$, in $(4.5)$ evaluated at the point  ${x^\mu}_i$ where the $i^{th}$ particle is found.
\par We then define a local connection form {\it in the neighborhood of the point $x_i$} as
$$ {\Gamma^\sigma}_{\lambda \gamma}(x_i) = {\partial {x^\sigma}_i \over \partial {\xi^\mu}_i} {\partial^2 {\xi^\mu}_i \over \partial{ x^\lambda}_i \partial {x^\gamma}_i} \eqno(4.12)$$
also, since it is a property of the manifold, as a {\it field} evaluated at the point  ${x^\mu}_i$, so that the geodesic equations can be written as
$$  {{\ddot x}^\sigma}_i = - {\Gamma^\sigma}_{\lambda \gamma}(x_i) {{\dot x}^\gamma}_i {{\dot x}^\lambda}_i
  - { 1\over M_i} g^{\sigma \lambda}(x_i) {\partial V(x_1, x_2,\dots x_N)\over\partial {x^\lambda}_i }. \eqno(4.13)$$
  \par The connection form in this case then also satisfies $(1.20)$ at each point $x_i$. Since this connection form coincides with Einstein's, the same method can be used to construct a Ricci tensor; the resulting Einstein equations will therefore have the same form, although there will necessarily be differences in the structure of the energy momentum tensor.\footnote{*}{Note that even in the absence of a potential function, the solutions for $g_{\mu\nu}(x)$ would reflect the many body structure of the energy momentum tensor through the Einstein equations.}   The empty space solution [36] is applicable in this framework as well, providing an interesting example for application [39](see also [40]), and the homogeneous case of Robertson, Friedman and Walker[41][42][43][44][45][46][47] would have a similar form to the well-known solution. Applications of this type will be investigated in succeeding papers. 
  \par Following the same procedure as for $(1.24)$, with general functions \hfil\break $F(x_1, x_2,\dots x_N,p_1, p_2,\dots p_N)$ with variables $\xi_1, \xi_2,\dots \xi_N, \pi_1, \pi_2,\dots \pi_N$ in the cotangent bundle assigned to the points $x_1, x_2,\dots x_N,p_1, p_2,\dots p_N$ in the general phase space for the $N$-body system, the Poisson bracket is defined by 
$$ \eqalign{{d F(\xi_1, \xi_2,\dots \xi_N, \pi_1, \pi_2,\dots \pi_N) \over d\tau} &= \Sigma_i\bigl({\partial F(\{\xi,\pi\})\over \partial{\xi^\mu}_i}{{\dot \xi}^\mu}_i +{\partial F(\{\xi,\pi\})\over \partial\pi_{\nu i}} {\dot \pi}_{\mu i}\bigr)\cr
&=\Sigma_i\bigl( {\partial F(\{\xi,\pi\})\over \partial{\xi^\mu}_i}{\partial K \over \partial \pi_{\mu i}} -{\partial F(\{\xi,\pi\})\over \partial\pi_{\mu i}} {\partial K \over \partial {\xi^\mu}_i}\bigr)\cr
&\equiv [F,K]_{PB} (\{\xi,\pi\}).\cr} \eqno(4.14) $$
The local transformations on the differentials cancel as for the one particle case (at each point $x_i$), so the Poisson bracket remains in the same form on the $8N$ dimensional phase space $\{x_i, p_i\}$ phase space.  We therefore have
$$\eqalign{{d F(x_1, x_2,\dots x_N,p_1, p_2,\dots p_N) \over d\tau}&=\Sigma_i{\partial F(\{x,p\})\over \partial{x^\mu}_i}{\partial K \over \partial p_{\mu i}} -{\partial F(\{x,p\})\over \partial p_{\mu i}} {\partial K \over \partial {x^\mu}_i}\cr 
&\equiv [F,K]_{PB} (\{x,p\}).\cr}\eqno(4.15)$$
In general, for two functions $A(\{x,p\})$ and $B(\{x,p\})$, the many body Poisson bracket is then
$$[A, B]_{PB} = \Sigma_i\bigl({\partial A(\{x,p\})\over \partial{x^\mu}_i}{\partial B(\{x,p\}) \over \partial p_{\mu i}} -{\partial A(\{x,p\})\over \partial p_{\mu i}} {\partial B(\{x,p\}) \over \partial {x^\mu}_i}\bigr). \eqno(4.16)$$
 Since the variables $x_1, x_2,\dots x_N,p_1, p_2,\dots p_N $ are to be considered as kinematically independent, we obtain the canonical bracket 
$$ [{x_i}^\mu ,p_{j\nu}]_{PB} = \delta_{ij} {\delta^\mu}_\nu \eqno(4.17)$$
\par We now turn to the equations of motion for $p_{\mu i}$. At the point ${x_i}^\mu $, as we have argued above,
$$ p_{\mu i} = {\partial {\xi^\lambda}_i \over \partial {x_i}^\mu}(x_i) \pi_{\lambda i}. \eqno(4.18))$$
It therefore follows, in the same way that we obtained $(1.33)$, that
$$\eqalign{{\dot p}_{\mu i} &=-{\partial V(x_1, x_2,\dots x_N,p_1)\over\partial {x^\mu}_i}+M_i g_{\sigma\gamma}(x_i){\Gamma^\sigma}_{\lambda \mu}(x_i) {{\dot x}^\gamma}_i {{\dot x}^\lambda}_i\cr
&= -{\partial V(x_1, x_2,\dots x_N,p_1)\over\partial {x^\mu}_i}+ {M_i \over 2}\bigl(g_{\sigma\gamma}(x_i){\Gamma^\sigma}_{\lambda \mu}(x_i)+g_{\sigma\lambda}(x_i){\Gamma^\sigma}_{\gamma \mu}(x_i)\bigr) {{\dot x}^\gamma}_i {{\dot x}^\lambda}_i.\cr} \eqno(4.19)$$
As for the one body case, this result also follows from the Lagrangian $(2.8)$.
\par The time rate of change of the canonical momentum is coupled, as for the geodesic motions of the  $x_i$, to the other $N-1$ particles through the potential function (and the energy momentum tensor).

\bigskip
\noindent{\it 5. Quantum Theory on the Curved Space}
\bigskip
\par The Poisson bracket formulas $(1.25)$ and $(1.26)$ can be considered as a basis for defining a quantum theory with canonical commutation relations
$$[x^\mu, p_\nu] = i\hbar{\delta^\mu}_\nu, \eqno(5.1)$$
so that\footnote{*} {As remarked above, $(5.1)$ also implies that $[x^\mu, p^\nu] = i\hbar g^{\mu\nu}(x)$, but application to polynomials in $p^\mu$ would introduce factors of $g^{\mu\nu}(x)$ and would require some care [31][32].} 
$$ [p_\mu, F(x)] = -i\hbar {\partial F \over \partial x^\mu}, \eqno(5.2)$$
and
$$[x^\mu, F(p)] = i\hbar {\partial F(p) \over \partial p_\mu}. \eqno(5.3)$$
The transcription of the Stueckelberg-Schr\"odinger equation for a wave function $\psi_\tau (x)$ can be taken to be (see also Schwinger and DeWitt [48][49][50][51])
$$i {\partial \over \partial \tau}\psi_\tau (x)= K \psi_\tau (x), \eqno(5.4)$$ 
where the operator valued Hamiltonian can be taken to be the Hermitian form,
on a Hilbert space defined with scalar product (with invariant measure; we write $g=-det\{g^{\mu\nu}\} $),
$$(\psi, \chi) = \int d^4 x \sqrt{g}{\psi^*}_\tau (x) \chi_\tau (x).
\eqno(5.5)$$
To construct a Hermitian Hamiltonian, we first study the properties of the canonical momentum in coordinate representation. Clearly, in coordinate representation, $-i{\partial \over \partial x^\mu}$ is not Hermitian due to the presence of the factor $\sqrt{g}$ in the integrand of the scalar product. The problem is somewhat analogous to that of Newton and Wigner [52]  in their treatment of the Klein Gordon equation in momentum space.  It is easily seen that the  operator
$$ p_\mu = -i {\partial \over \partial x^\mu} - {i \over 2}{1 \over \sqrt{g(x)}}{\partial \over \partial x^\mu} \sqrt{g(x)}  \eqno(5.6)$$
is essentially self-adjoint in the scalar product $(5.5)$, satisfying as well the commutation relations $(5.1)$.\footnote{*}{The physically observable momentum can be defined, as in $(1.15)$,  as ${1\over 2} \{g^{\mu\nu},p_\nu\}$, with commutation relations similar of the form $(1.27)$.}
\par Since $p_\mu$ is Hermitian in the scalar product $(5.6)$, we can write the Hermitian Hamiltonian as
$$K = {1 \over 2M}p_\mu g^{\mu \nu}  p_\nu  + V(x), \eqno(5.7)$$
consistent with the local coordinate transformation of $(1.1)$.
\par The normalization condition over the manifold $\{x\}$ is not a trivial transcription of the Euclidean condition on the SHP quantum theory [1]. If we think of the integral $(5.5)$ as constructed from integrating over coordinate components, a large excursion along a coordinate of the curved space may bring one, perhaps many times, to a nearby (Euclidean) neighborhood of some point. The integration $(5.5)$ must be considered as a total volume sum with invariant measure on the whole space, consistent with the notion of Lesbesgue measure and the idea that the norm is the sum of probability measures on every subset contained. The procedure for carrying out such integrals would, of course, depend on the geometrical structure of the manifold.
\par This construction can be carried over to the many body case directly, {\it i.e}, with the operator properties of the coordinates and momenta
$$[{x^\mu}_i, p_{\nu i}] = i\hbar{\delta^i}_j{\delta^\mu}_\nu, \eqno(5.8)$$
and therefore
$$ [p_{\mu i}, F(\{x\})] = -i\hbar {\partial F(\{x\}) \over \partial {x^\mu}_i}, \eqno(5.9)$$
and
$$[{x^\mu}_i, F(\{p)\}] = i\hbar {\partial F(\{p\}) \over \partial p_{\mu i}}. \eqno(5.10)$$
The scalar product is then (the flat space Lorentz invariant $d^4x$ goes over to the local diffeomorphism invariant $d^4x \sqrt{g}$)
$$(\psi, \chi) = \int \Pi_i \big\{d^4(x_i)\sqrt{g(x_i)}\big\}{\psi^*}_\tau (x_1, x_2,\dots x_N) \chi_\tau (x_1, x_2,\dots x_N). \eqno(5.11)$$
In this scalar product, the Hamiltonian (with $(5.6)$ for each $p_{\mu i}$ at ${x^\mu}_i$) 
$$K = \Sigma_i {1 \over 2M_i}p_{\mu i} g^{\mu \nu}(x_i)  p_{\nu i}  + V(x_1, x_2,\dots x_N) \eqno(5.12)$$
is essentially self-adjoint.  
\bigskip
\bigskip{\it 6. Fourier Transform, Potential Scattering Theory and the Propagator }
\bigskip
\par In the context of quantum field theory and gravitons, Bjerrum-Bohr {\it et al} [16] have discussed scattering theory to arrive at aspects of classical general relativity, providing interesting motivation for a scattering theory in general relativity. In this section, we develop a potential scattering theory in the framework of the quantum theory we have described in the previous section. In case the potential $V$ is zero  (or constant), we also discuss how the ``free'' particle propagator is affected by the curvature of the manifold. 
\par To deal with this problem, we discuss first the formulation of the Fourier transform $f(x) \rightarrow {\tilde f}(p)$ for a scalar function $f(x)$ (we shall use $x^\mu$ and the canonically conjugate $p_\mu$ in this discussion). Let us define ($g \equiv -\det g_{\mu\nu}$)
$$ {\tilde f}(p) = \int d^4x \sqrt{g(x)} e^{ip_\mu x^\mu} f(x). \eqno(6.1)$$
   The inverse is given by
$$ \eqalign{ \int e^{-ip_\mu x^\mu}{\tilde f}(p) d^4 p &= \int d^4p e^{-ip_\mu(x^\mu -{x'}^\mu)}f(x') \sqrt{g(x')}d^4x'\cr
&= (2\pi)^4 f(x) \sqrt{g(x)} \cr} \eqno(6.2)$$
so  that\footnote{*}{A simple but nontrivial proof of this result will be given elsewhere.}     
$$f(x) = {1 \over (2\pi)^4 \sqrt{g(x)}}  \int e^{-ip_\mu x^\mu}{\tilde f}(p) d^4 p.  \eqno(6.3)$$
\par One sees immediately that under diffeomorphisms, for which with the scalar property $f(x) = f'(x')$ ,  ${\tilde f}(p) \rightarrow {\tilde f}'(p)$ . 
The Fourier transform of $f'(x')$ we define as
$$ {\tilde f'}(p) = \int d^4x' \sqrt{g(x')} e^{ip_\mu {x'}^\mu} f'(x'), \eqno(6.4)$$
By change of integration variables, we have
$$ {\tilde f'}(p) = \int d^4x \sqrt{g(x)} e^{ip_\mu x^\mu} f'(x), \eqno(6.5)$$
\par  In Dirac notation,
$$ f(x) = <x|f>, \eqno(6.6)$$
and we write as well
$$ {\tilde f}(p) = <p|f>.\eqno(6.7)$$
For
$$ \eqalign{<x|p> &=   {1 \over (2\pi)^4 \sqrt{g(x)}}  e^{-ip_\mu x^\mu}\cr
<p|x> &= \sqrt{g(x)} e^{ip_\mu x^\mu}, \cr}\eqno(6.8)$$
we have, {\it e.g.}, the usual action of transformation functions
$$ \int <x|p><p|f>d^4 p = <x|f>, \eqno(6.9)$$
where we have used
$$\eqalign{ \int <x|p><p|x'> d^4p &= {1 \over (2\pi)^4 \sqrt{g(x)}}\int d^4p e^{-ip_\mu x^\mu}e^{ip_\mu {x'}^\mu}\sqrt{g(x')}\cr
&= \delta^4 (x-x'). \cr} \eqno(6.10)$$
Note that the the tranformation functions $<x|p>$ and $<p|x>$ are not simple complex conjugates of each other, but require nontrivial factors of $\sqrt{g(x)}$ and its inverse to satisfy the necessary transformation laws on the manifold.   
Conversely, (the factors $\sqrt{g(x)}$ and its inverse cancel)           
$$\int <p' |x><x|p> d^4x = \delta^4(p'-p). \eqno(6.11)$$
\par The formulation we have given above is explicitly covariant, and will be used in the sequel to discuss the scattering theory.  However, to demonstrate the role of the momentum operator as the generator of translations, we study an alternative representation.  
\par  The presence of $\sqrt{g(x)}$ in the integrations over $d^4 x$ is analogous to the Newton-Wigner discussion [52] in momentum space, where $d^3p/p_0$ is the Lorentz invariant measure for the Klein-Gordon scalar product; in our case, we are concerned with local diffeomorphism invarance. We consider, in analogy to the Newton-Wigner construction, the transformation from elements of the original Hilbert space, say $\psi(x)$, to a new representation, which we shall call the Newton-Wigner representation, 
$$ \psi_{NW}(x) = (g(x))^{ {1\over 4}}\psi (x). \eqno(6.12)$$
In this representation, the norm
$$ \int d^4x |\psi_{NW}(x)|^2 = \int d^4x \sqrt{g(x)} |\psi(x)|^2 \eqno(6.13)$$
is conserved, implying equivalence.
\par A locally defined operator ${\cal O}$ acting on $\psi(x)$ as
$$ {\cal O}\psi(x) = {\cal O}(g(x))^{- {1\over 4}}\psi'(x) \eqno(6.14)$$
implies that operators ${\cal O}$ should be replaced by
$$ {\cal O}'= (g(x))^{ {1\over 4}}{\cal O}(g(x))^{- {1\over 4}} \eqno(6.15)$$
Applying this transformation (a scalar analog, as in [52], to  the Foldy-Wouthuysen transformation [53])to the canonical momentum operator, one finds
$$ (g(x))^{ {1\over 4}}\bigl(-i {\partial \over \partial x^\mu} - {i \over 2}{1 \over\sqrt{g(x)}} {\partial \over \partial x^\mu} \sqrt{g(x)}\bigr)(g(x))^{- {1\over 4}}
= -i {\partial \over \partial x^\mu}, \eqno(6.16)$$
providing a simple representation of the generator of translations. As for the Newton-Wigner construction in momentum space, where explicit Lorentz covariance would be obscured, this transformation obscures the local diffeomorphism covariance of the theory. 
\par The property of translation for the operator $p_\mu$ can be clearly demonstrated in the Newton-Wigner representation $(6.12)$. We now have 
$$ \eqalign{{\tilde\psi}(p) \equiv <p|\psi_{NW}> &= \int d^4x \sqrt{g(x)} e^{ip_\mu x^\mu} (g(x))^{-{1 \over 4}} \psi_{NW} (x) \cr
&= \int d^4x e^{ip_\mu x^\mu} (g(x))^{1 \over 4} \psi_{NW} (x).\cr} \eqno(6.17)$$ 
The inverse is then
$$ \psi_{NW} (x) \equiv <x|\psi_{NW}> = {1 \over (2\pi)^4 (g(x))^{1\over 4}} \int d^4p e^{-p_\mu x^\mu} {\tilde\psi}(p). \eqno(6.18)$$
We then identify
$$<p|x>_{NW} = g(x)^{1 \over 4} e^{ip_\mu x^\mu} \eqno(6.19) $$
and
$$<x|p>_{NW} =  {1 \over (2\pi)^4 g(x)^{1\over 4}}e^{-ip_\mu x^\mu}  \eqno(6.20)$$
As for the covariant form discussed above, we have 
$$ \int <p'|x>_{NW} <x|p>_{NW} d^4x = \delta^4(p-p'), \eqno(6.21)$$
and
$$ \int <x|p>_{NW}<p|x'>_{NW} d^4p = \delta^4 (x-x'). \eqno(6.22)$$
\par Using these transformation functions to bring the operator $p_\lambda$ defined in $(5.6)$ to momentum representation, one obtains, for $(p_\lambda)_{op}$ the expression $(5.6)$, with $(6.16)$,
$$ \eqalign {\int d^4x <p|x>_{NW} (p_\lambda)_{op}<x|p'>_{NW} &= {1 \over (2\pi)^4}\int d^4x
g(x)^{ 1\over 4}   (p_\lambda)_{op} { 1 \over g(x)^{1\over 4}} e^{i{p_\mu -{p'}_\mu }x^\mu}\cr
&= {1 \over (2\pi)^4}\int d^4x
\bigl(-i {\partial\over\partial x^\lambda}\bigr)  e^{i(p_\mu -{p'}_\mu )x^\mu}\cr
&= p_\lambda \delta^4(p-p'),\cr}.\eqno(6.23)$$
so that in the Newton-Wigner representation, the operator $(5.6)$ becomes translation (along the coordinate curves).
\par In the following, we maintain the explicitly covariant form of the theory.
\par We may now formulate the potential scattering problem in interaction picture. Let us write for $(4.7)$
$$ K= K_0 + V \eqno(6.24)$$
where, for example, for the two body problem\footnote{*}{The result here can be immediately extended to the $N$-body problem.}, $V=V(x_1,x_2)$ (a scalar function for diffeomorphisms at $x_1$ and $x_2$ for all $x_1$ and $x_2$) and (for the self-ajoint $p_\mu$ in $x$ representation)
$$ K_0 ={1 \over 2M_1} {p_\mu}^1  g^{\mu\nu}(x_1){p_\nu}^1+{1 \over 2M_2} {p_\mu}^2 g^{\mu\nu}(x_2){p_\nu}^2. \eqno(6.25)$$
\par In the scattering of large masses, the metric $g^{\mu\nu}$ may depend on $\tau$. In an adiabatic sense, we shall assume here that there is no explicit $\tau$ dependence in the  metric. 
\par If we then write for the two body wave function
$$ \psi_\tau (x_1,x_2) = e^{-iK_0\tau} \chi_\tau (x_1,x_2), \eqno(6.26)$$
the Stueckelberg-Schr\"odinger equation for $\chi_\tau$ becomes, as for the usual interaction picture,
$$ i {\partial  \chi_\tau (x_1,x_2) \over \partial \tau} = V(x_1,x_2, \tau)
\chi_\tau(x_1,x_2), \eqno(6.27)$$
where
$$V(x_1,x_2, \tau)= e^{iK_0\tau}V(x_1, x_2)e^{-iK_0\tau}. \eqno(6.28)$$
At $\tau\rightarrow -\infty$, we shall assume that the wave function moves out of the region of spacetime where the two body potential is effective\footnote{**}{In a model in which the potential term $V$ represents ``dark energy''[19][20], modulated by the proximity of, in this case, two massive systems, this limit would correspond to a configuration in which the mutual influence of the two systems becomes negligible and each is in an environment where the dark energy is distributed, as $V_1 \oplus V_2$, for example, in accordance with the requirements of MOND[20][21][22][23][24][25][26][27]. }, so that we have a relation that can be studied, as in usual scattering theory, by iteration:
$$ \chi_\tau (x_1, x_2) = \chi_{-\infty} (x_1, x_2) -i \int_{-\infty}^\tau d\tau' V(x_1,x_2,\tau') \chi_{\tau'}(x_1,x_2). \eqno(6.29))$$
\par It then follows that $\chi_\tau (x_1, x_2)$ can be expressed as a $\tau$ ordered product
$$ \chi_\tau (x_1, x_2) = \bigl(e^{\int_{-\infty}^\tau -iV(\tau')d\tau'}\bigr)_+ \chi_{-\infty} (x_1, x_2). \eqno(6.30)$$ 
Now, recall the usual scattering condition. Defining $U(\tau) = e^{-iK\tau}$ and $U_0(\tau) = e^{-iK_0\tau}$, we assume that for a state $\psi$ that evolves to $\psi_\tau$ by $U(\tau) = e^{-iK\tau}$ that there is an {\it in} state for which
$$ \Vert U(\tau) \psi - U_0(\tau)\psi_{in}\Vert \rightarrow 0 \ \ \ \tau\rightarrow-\infty \eqno(6.31)$$
and 
$$ \Vert U(\tau) \psi - U_0(\tau)\psi_{out}\Vert \rightarrow 0 \ \ \ \tau\rightarrow +\infty, \eqno(6.32)$$
which, for sufficiently dense $\psi_{in}$ and $\psi_{out}$ define the wave operators
$\Omega+$ and $\Omega_-$ respectively as
$$\eqalign{\psi &= U^\dagger(\tau) U(0)\psi_{out}\rightarrow \Omega_- \psi_{out}\ \ \tau\rightarrow +\infty \cr
&= U^\dagger(\tau) U(0)\psi_{out}\rightarrow \Omega_+ \psi_{in}\ \ \tau\rightarrow -\infty.\cr} \eqno(6.33)$$
\par A sufficient condition for the existence of this limit for, {\it e.g.}, $\Omega_+$, is that $\Vert V U_0 \psi_{in}\Vert$ vanishes sufficiently rapidly (there are weaker conditions that may apply) so that it is integrable from $-\infty$ to zero in $\tau$. This conditon must be investigated for any particular model, but we shall assume here that it is satisfied.
\par It follows that
$$ \Omega_- \psi_{out} = \Omega _+ \psi_{in}, \eqno(6.34)$$
and then, by unitarity,
$$ \psi_{out} = {\Omega_-}^\dagger \Omega_+ \psi_{in} \equiv S \psi_{in}, \eqno(6.35)$$
defining the $S$-matrix.
\par Now, for
$$\chi_\tau = {U_0}(\tau)^\dagger \psi_\tau = {U_0}(\tau)^\dagger U(\tau) \psi, \eqno(6.36)$$
we see that
$$\eqalign{ \chi_{-\infty} &= {\Omega_+}^\dagger \psi\cr
            \chi_{+\infty} &= {\Omega_-}^\dagger \psi,\cr} \eqno(6.37)$$
so that
$$\psi = \Omega_+ \chi_{-\infty} = \Omega_- \chi_{+\infty}, \eqno(6.38)$$
or
$$ \chi_{+\infty}= S \chi_{-\infty}. \eqno(6.39)$$
Therefore $(6.30)$ provides, for $\tau \rightarrow \infty$, a formula for the $S$ matrix, as in standard scattering theory.
\par We now study the first few terms of the iteration of $(6.29)$. Consider
$$\eqalign{\chi_\tau (x_1,x_2) &=\chi_{in} -i \int_{-\infty}^\tau d\tau' V(\tau') \chi_{in}(x_1,x_2)\cr
&+ (-i)^2 \int_{-\infty}^\tau d\tau'\int_{-\infty}^{\tau'} d\tau'' V(\tau') V(\tau'')\chi_{in}(x_1,x_2)\cr
&+ (-i)^3 \int_{-\infty}^\tau d\tau' \int_{-\infty}^{\tau'} d\tau''\int_{-\infty}^{\tau''} d\tau''' V(\tau')V(\tau'')V(\tau''')\chi_{in}(x_1,x_2) +\cdots \cr} \eqno(6.40)$$
Inserting intermediate states corresponding to the spectrum of $p$, we see that we have terms of the type 
$$ \eqalign{<p_1p_2|V(\tau)|{ p_1}' {p_2}'> &=<p_1p_2|e^{iK_0 \tau} V e^{-iK_0\tau}|{ p_1}' {p_2}'> \cr
&= \int d{p_1}''d{p_2}''d{p_1}''' d{p_2}''' <p_1p_2|e^{iK_0 \tau}|{p_1}'' {p_2}''>\cr
&\times <{p_1}'' {p_2}''| V|{p_1}''' {p_2}'''>
<{p_1}''' {p_2}'''| e^{-iK_0\tau} |{ p_1}' {p_2}'>.\cr} \eqno(6.41)$$
Introducing the $x$ representation in intermediate states (through the $<x|p>$ transformation functions) is not useful here since $V$ is not necessarily Poincar\'e invariant. Moreover, $K_0$ is not diagonalized in the $p$ representation, so that the $\tau$ integrations cannot be carried out leading to Feynman propagators in the usual way [54][55][56]. However, we proceed as follows.
\par The two terms of $K_0$,
$$ K_0 = K_{01} + K_{02}= {1 \over 2M_1} p_{1\mu} g^{\mu\nu}(x_1) p_{1\nu} +  {1 \over 2M_2} p_{2\mu} g^{\mu\nu}(x_2) p_{2\nu} \eqno(6.42)$$
commute, and therefore
$$<p_1p_2|e^{iK_0 \tau}|{ p_1}' {p_2}'>= <p_1|e^{iK_{01} \tau}|{ p_1}' ><p_2|e^{iK_{02} \tau}| {p_2}'>\eqno(6.43)$$
\par Let us consider the matrix element of $K_{01}$ in momentum eigenstates
$$\eqalign{<p_1|K_{01} |{p_1}'> &= \int d^4 x_1 \sqrt{g(x_1)} {1 \over 2M_1}p_{1\mu} {p_{1\mu}}'<p_1|x_1> g^{\mu\nu}(x_1)<x_1|{ p_1}'>\cr
&= {1 \over 2M_1}p_{1\mu} {p_{1\mu}}'\int d^4x_1\sqrt{g(x_1)}e^{ip_{1\kappa} {x_1}^\kappa} g^{\mu\nu}(x_1)
{1\over (2\pi)^4 \sqrt{g(x_1)}} e^{-i{{p_1}'}^\sigma x_{1\sigma}} \cr
&= {1 \over 2M_1}p_{1\mu} {p_{1\mu}}'{\tilde g}_{\mu\nu}(p_1 -{ p_1}'), \cr},\eqno(6.44)$$
where ${\tilde g}^{\mu\nu}(p)$ is the Fourier transform of $ g^{\mu\nu}(x)$.
\par In coordinate representation, each of the $K_{0i}$ is a (essentially self-adjoint) Laplace Beltrami operator  (see, for example, S. Helgason[57]) and has, in general, continuous spectrum in $(-\infty, \infty)$. Since the  two $K_{0i}$ commute, the operator $K_0$ defined in $(6.42)$ is a direct sum on the product Hilbert space. Suppose its generalized eigenfunctions are $\{|\lambda_1,\lambda_2>\}$, so that
$$ K_0|\lambda_1,\lambda_2>=( \lambda_1 +\lambda_2)|\lambda_1,\lambda_2>. \eqno(6.45)$$ 
In the following, we call
$$ |\lambda_1 \lambda_2> \equiv |\lambda^{(2)}>. \eqno(6.46)$$
\par We now return to $(6.40)$ and introduce the complete set $\{|\lambda^{(2)}>\}$ as intermediate states:
$$\eqalign{<p_1p_2|\chi_\tau> &=<p_1p_2|\chi_{in}> -i \int_{-\infty}^\tau d\tau' <p_1p_2|V(\tau') \chi_{in}>\cr
&+ (-i)^2 \int_{-\infty}^\tau d\tau'\int_{-\infty}^{\tau'} d\tau''\int d\lambda^{(2)}<p_1p_2| V(\tau')|\lambda^{(2)}><\lambda^{(2)}| V(\tau'')\chi_{in}>\cr
&+ (-i)^3 \int_{-\infty}^\tau d\tau' \int_{-\infty}^{\tau'} d\tau''\cr
&\int_{-\infty}^{\tau''} d\tau''' \int d\lambda^{(2)} d\lambda^{(2)'} <p_1p_2|V(\tau')|\lambda^{(2)}>\cr
&<\lambda^{(2)}|V(\tau'')|\lambda^{(2)'}><\lambda^{(2)'}
|V(\tau''')\chi_{in}> +\cdots \cr} \eqno(6.47))$$
Let us consider the third order term. Inserting the transformation function $<p_1p_2|\lambda^{(2)''}>$ for the first factor to make explicit the matrix element $<\lambda^{(2)''}|V(\tau')|\lambda^{(2)}>$ and in the last factor $<\lambda^{(2)'''}|{p_1}'{p_2}'>$ to obtain the matrix element of $V(\tau''')$ in the $\lambda^{(2)}$  representation, for which
$$<\lambda^{(2)''}|V(\tau')|\lambda^{(2)}>= e^{i({\lambda_1}'' +{\lambda_2}''-\lambda_1 -\lambda_2)\tau'}<\lambda^{(2)''}|V|\lambda^{(2)}>, \eqno(6.48)$$
we may carry out the $\tau$ integrations to obtain
$$\eqalign{&(-1)^3 \int d\lambda^{(2)}d\lambda^{(2)'}d\lambda^{(2)''}d\lambda^{(2)'''}d^4{p_1}' d^4 {p_2}'
<p_1 p_2|\lambda^{(2)}>\cr
&{<\lambda^{(2)}|V|\lambda^{(2)'}> <\lambda^{(2)'}|V|\lambda^{(2)''}>
 <\lambda^{(2)''}|V|\lambda^{(2)'''}>     \over (\lambda_1 +\lambda_2 - {\lambda_1}^{'''}-{\lambda_2}^{'''}-i\epsilon)
({\lambda_1}' +{\lambda_2}' - {\lambda_1}^{'''}-{\lambda_2}^{'''}-i\epsilon)({\lambda_1}'' +{\lambda_2}'' - {\lambda_1}^{'''}-{\lambda_2}^{'''}-i\epsilon)}\cr
&\times <\lambda^{(2)'''}|{p_1}'{p_2}'><{p_1}'{p_2}'|\chi_{in}>. \cr}  \eqno(6.49)$$
We have inserted a factor $e^{\epsilon \tau}$ for convergence as $\tau \rightarrow -\infty$ in the last factor. This  procedure is based in the flat space limit on $\Vert Ve^{-iK_0 \tau} \psi \Vert$ vanishing as $\tau \rightarrow -\infty$; if our asymptotic condition is in flat space, the potential term would have the same features as in the SHP theory [1], for which this condition can hold. The $\tau$ integrations, starting from the last, carry over a factor of $e^{-i\epsilon \tau} $ to each successive integration, providing the $-i\epsilon$ terms in the denominators, as in the usual scattering theory.  The same general structure, with alternating signs, obtains to every order.
\par The structure of the intermediate propagators is similar to the usual (two-body) Feynman free propagators, but the evaluation of the vertices $<\lambda^{(2)}| V|\lambda^{(2)'}>$ involves the transformation functions $<\lambda^{(2)}|x_1,x_2>$, known from solutions for the Laplace-Beltrami spectral problem. The $\lambda $'s play the same role here as the energy eigenvalues (continuous spectrum) for the unperturbed Hamiltonian in the nonrelativistic scattering theory, but in this case they correspond to the spectrum of the Laplace-Beltrami operators that constitute the Hamiltonian for evolution on the manifold. 
\par It is interesting to compare this result with the emergence of the propagator from the Green's function for a single particle (or, in a simple generalization, to many particle) propagation. To see this, let us compute the evolution of a ``free'' one body state (here $K_0$ has just one term of $(6.42)$)
$$ {\psi^0}_\tau (p) = \int d^4p'  <p|e^{-iK_0 \tau} |p'> {\psi^0}_0(p')  \eqno(6.50)$$
The Laplace transform on $(0,\infty)$ for $Im\ \ s>0$ is  then
$$ {{\tilde \psi}_s}^0 = i \int d^4 p' < p| {1 \over s-K_0}|p'>{\psi^0}_0(p'), \eqno(6.51)$$
defining the ``free'' one particle propagator as\footnote{*}{In the $\{\lambda\}$ representation, for free evolution, this formula would provide the same denominator as occurs in $(6.49)$.}
$$ G(s) = { 1 \over s- K_0}. \eqno(6.52)$$
For an almost flat space where ($\eta^{\mu\nu} = diag(-,+,+,+)$)
$$ g^{\mu\nu}(x) = \eta^{\mu\nu} + h^{\mu\nu}(x), \eqno(6.53)$$
the bilinear perturbation term corresponds to a Laplace-Beltrami type mass operator in addition to the value of $p^\mu p_\mu = -m^2$, {\it i.e.},
$$G(s) \cong { 1 \over s- {p^\mu p_\mu\over 2M}- {1\over 2M}p_\mu h^{\mu\nu}(x) p_\nu} \eqno(6.54)$$
\par We now turn to discuss the introduction of electromagnetism, for a single particle and for many particle systems.
\bigskip
\noindent{\it 7. Electromagnetism}
\bigskip
\par As C.N. Yang [58] wrote, electromagnetism can be thought of as a $U(1)$ fiber bundle. The electromagnetic potential vector field emerges as a section on the fiber bundle in the gauge transformations of the quantum theory. To illustrate this idea, consider 
what happens to Eq. $(5.4)$ if we consider, instead of the function $\psi_\tau(x)$, the function ${\psi'}_\tau(x)=e^{i\Lambda(x,\tau)}\psi_\tau(x) $ resulting from a unitary transformation  $e^{i\Lambda(x,\tau)}$ defined locally (for $\Lambda(x,\tau)$ a scalar function) on the Hilbert space at each value of $\tau$. Since $p_\mu$ acts like a derivative on $x^\mu$, it differentiates $\Lambda(x,\tau)$, just as for the corresponding computation in the flat Minkowski space. As for the flat space case, we must add a gauge compensation term so that
$$ (p_\mu -a'_\mu(x,\tau))e^{i\Lambda(x,\tau)}\psi_\tau(x) =e^{i\Lambda(x,\tau)}(p_\mu -a_\mu(x,\tau))\psi_\tau(x), \eqno(7.1)$$
{\it i.e.}, assuring that $(p_\mu -a_\mu(x,\tau))\psi_\tau(x)$ is an element of the Hilbert space, transformed locally at every point in the same way, and therefore undergoes the same unitary transformation as $ \psi_\tau(x)$.
Carrying out the derivative implied by the action of $p_\mu$ (as in $(5.6)$), we find the condition that
$$ a'_\mu(x,\tau) = a_\mu(x,\tau) + {\partial \Lambda(x,\tau) \over \partial x^\mu}, \eqno(7.2)$$
the usual form of a gauge transformation. From the scalar nature of the wave function, we have implicitly assumed Einstein's property of general covariance for the fields $a_\mu(x,\tau)$.
\par Unless we restrict ourselves to the so-called ``Hamilton gauge'' (with $\Lambda$ independent of $\tau$), the form of $(7.4)$ implies the existence of a fifth field [1][59][60] $a_5(x,\tau)$, for which we must have
$$ \bigl\{i {\partial \over \partial \tau} + a'_5(x,\tau)\bigr\}\psi'_\tau (x)
= e^{i\Lambda(x,\tau)} \bigl\{i {\partial \over \partial \tau} + a_5(x,\tau)\bigr\}\psi_\tau (x)
\eqno(7.3)$$
By the same argument, we then have
$$ a'_5(x,\tau) = a_5(x,\tau) + {\partial \over \partial \tau} \Lambda(x,\tau). \eqno(7.4)$$
The Stueckelberg-Schr\"odinger equation then becomes
$$ i {\partial \over \partial \tau}\psi_\tau (x) =\big\{{1 \over 2M}(p_\mu-a_\mu(x,\tau)) g^{\mu \nu} ( p_\nu-a_\nu(x,\tau)) -a_5(x,\tau)(x)\big\}\psi_\tau(x), \eqno(7.5)$$
where the scalar field of the potential model is now replaced by the generally  $\tau$ dependent $a_5(x,\tau)$.
\par In the usual way, we can define in the flat tangent space, a gauge invariant field strength[1][59][60]
$$ {\tilde f}_{\alpha \beta}(\xi, \tau) = \partial_\alpha a_\beta (\xi, \tau) - \partial_\beta a_\alpha (\xi,\tau),
\eqno(7.6)$$
where $\alpha, \beta = (0,1,2,3,5)$, which satifies the equation
$$ \partial^\alpha {\tilde f}_{\alpha \beta}(\xi, \tau)= j_\beta(\xi,\tau). \eqno(7.7)$$
\par The first four components of the current have the form of a Jackson type construction [61] before integration over $\tau$, and the fifth  component is the density $\rho(\xi,\tau) \propto {\psi^*}_\tau(\xi)\psi_\tau(\xi)$ in the SHP theory (see [1] for details).
\par It is easy to see that a coordinate transformation leads to the rule of replacement of derivatives by covariant derivatives so that in the curved space
$$\eqalign{ f_{\mu \nu}(x, \tau) &= a_{\mu ; \nu} - a_{\nu ; \mu} = {\partial {\xi^\sigma} \over \partial {x^\nu}}{\partial {\xi^\lambda} \over \partial {x^\mu}}{\tilde f}_{\lambda\sigma}\cr
f_{\mu 5} &= \partial_\mu a_5 - \partial_5 a_\mu ,\cr} \eqno(7.8)$$
since $a_5$ is a Lorentz scalar.
\par For the fourth and fifth components, we have
$$ \eqalign{ f_{\mu \nu}^{;\mu}(x, \tau) + \partial^5 f_{5 \nu}(x, \tau) &= j_\nu(x,\tau);\cr
{ f_{\mu 5}}^{;\mu}(x,\tau) &= j_5 (x,\tau) = \rho(x,\tau),\cr} \eqno(7.9)$$
where the last is analogous to the non-relativistic $\nabla \cdot {\bf E} =\rho$. Clearly, the covariant divergence of $j_\nu(x,\tau)$ vanishes.
\par We now study the structure of the corresponding current. To do this, we  write an action
for which the variation with respect to ${\psi^*}_\tau(x)$ yields the Stueckelberg-Schr\"odinger equation $(7.5)$,
$$\eqalign{S &=\int d\tau d^4x \sqrt{g} \bigl\{ i {\psi^*}_\tau(x) {\partial \over \partial \tau} \psi_\tau(x) -i \psi_\tau(x) {\partial \over \partial \tau} {\psi^*}_\tau(x)
+ a_5(x, \tau){\psi^*}_\tau(x)\psi_\tau(x) \cr
&-\bigl\{  {\psi^*}_\tau(x) \big\{{1 \over 2M}(p_\mu-a_\mu(x,\tau)) g^{\mu \nu} ( p_\nu-a_\nu(x,\tau)) -a_5(x,\tau)\big\}\psi_\tau(x) \cr }\eqno(7.10)$$
where $f^{\mu\nu} (x,\tau) = g^{\mu \lambda}g^{\nu \sigma} f_{\lambda \sigma} (x,\tau)$.
We add to the action a purely electromagnetic part
$$ S_{em}= +  {1 \over 4\sqrt{g}}\bigl( f^{\mu \nu} f_{\mu \nu}
+ f^{\mu 5}f_{\mu 5}\bigr), \eqno(7.11)$$
where (since $a_5$ is scalar its covariant derivative is an ordinary derivative)
$$ f_{\mu 5} = \partial_\mu a_5 - \partial_\tau a_\mu \eqno(7.12)$$
and $a^5 = g^{5\alpha} a_\alpha$  depending on the metric for the embedding of $O(4,1)$ or $O(3,2)$ chosen for the $5D$ manifold. As for the nonrelativistic theory on $3D$, where the gauge fields make accessible the $(3,1)$ manifold of Minkowski space, the gauge fields of the $(3+1)D$ theory make accessible the embedding of the  $(4,1)$ or $(3,2)$ manifold. As we shall see, however,  in our discussion of the many body problem, the assumption of universality in $\tau$ does not admit such a higher symmetry.
\par In 1995, Land, Shnerb and Horwitz [60] studied the consequences of assuming covariant commutation relations between $x^\mu$ and ${\dot x}^\nu$ on a manifold using a theorem of Hojman and Sheply [62]  extending and generalizing the work of Tanimura [63].  Their results, including the development of the 5D theory, agree in the one particle sector with what we have presented here. 
\par To fully treat such a development with the methods we have used here, one would have to start with a one degree higher dimensional Stueckelberg equation; its gauge fields would open the possibility of a $6D$ manifold as a result of gauge invariance.  We shall, however,  truncate this sequence here at the level of $4D$, retaining $\tau$ as the universal invariant parameter of evolution.
\par We now obtain the current by variation of $a_\mu$ in the action. Integrating by parts in the kinetic term (for the self-adjoint $p_\mu = -i {\partial \over \partial x^\mu}-{i\over 2}{1\over \sqrt{g(x)}}{\partial \over \partial x^\mu}\sqrt{g(x)}$), we have
$$\eqalign{S_{kin} &= - {1 \over 2M}\int d\tau d^4x\sqrt{g(x)} ((p_\mu -a_\mu)\psi)^* g^{\mu\nu}(p_\nu -a_\nu)\psi\cr
&= +{1 \over 2M}\int d\tau d^4x\sqrt{g(x)} ((p_\mu +a_\mu)\psi^*) g^{\mu\nu}(p_\nu -a_\nu)\psi,\cr} \eqno(7.13)$$
so that
$$ {\delta S_{kin} \over \delta a_\mu} =  {\sqrt{g(x)} \over 2M} \bigl(\psi^* g^{\mu\nu}(p_\nu -a_\nu)\psi - ((p_\nu +a_\nu)\psi^*)g^{\mu\nu}\psi \bigr). \eqno(7.14)$$
 Here, $ \psi^* \psi (x,\tau)$ is the probability to find the particle (event) in the invariant volume element $ \sqrt{g}d^4x$, so that $\psi^* \psi (x,\tau)$  must go over to $ {1 \over \sqrt{g(x)}}\delta^4(x-x')$ in the classical limit (see Weinberg [37]). Therefore, we must define the current as\footnote{*}{Note that if we follow the method of Jackson[61], defining the macroscopic current
$$J^\mu(x) =  {1 \over \sqrt{g(x)}}\int d\tau {\dot x}^\mu \delta^4(x-x(\tau)),$$
then $\partial_\mu  \sqrt{g(x)}J^\mu) = 0$.}
$$ j_\mu(x,\tau) = {1 \over \sqrt{g(x)}} {1 \over 2M} \bigl(\psi^* g^{\mu\nu}(p_\nu -a_\nu)\psi - ((p_\nu +a_\nu)\psi^*)g^{\mu\nu}\psi \bigr), \eqno(7.15)$$
in agreement in form with the corresponding known nonrelativistic formula ($g \rightarrow 1$ in the nonrelativistic limit, so the extra term in $p_\mu$ vanishes). Note further that the integral of the current over a hypersurface with the invariant measure $d^4 x \sqrt{g}$ has well-defined physical meaning.  
\par We now study the variation of the action with respect to  $a_5$. 
The variation of the full action (both $S_m$ and $S_{em}$) with respect to $a_5$ then yields the field equation
$$ {f_{\mu 5}}^{;\mu}(x,\tau) \equiv  {1 \over \sqrt{g(x)}} \rho(x,\tau)= {\psi_\tau}^*\psi_\tau(x,\tau) , \eqno(7.16)$$
in analogy to the standard Maxwell equation $\nabla \cdot {\bf E}= \rho$.
\par Furthermore, the variation with respect to $a_\mu$, with the definition of the current $(7.15)$, yields the covariant field equations (as one would conclude from the application of general covariance [64]) The variation of $a_\mu$ in $(7.10),\ (7.11)$ leads to   
$$ {f_{\mu \nu}}^{; \mu} (x, \tau) = j_\nu (x,\tau),  \eqno (7.17)$$
or, equivalently,
$$\partial_\mu (\sqrt{g} f^{\mu\nu}) =  j^\nu \eqno(7.18)$$
\par The Lorentz force (see also [59]) follows by directly transcribing the flat space formula (for charge unity),
$$ F^\mu= {f^\mu}_\nu {dx^\nu \over d\tau} \eqno(7.19)$$
\bigskip
\noindent{\it 8 The Many Body Problem for Electromagnetism}
\bigskip
\par In the following, we generalize this structure to the many body problem.
\par The many-body wave function can be written as the span of the direct product of wave functions associated with isomorphic one particle Hilbert spaces (which also may be used in the construction of the Fock space on the manifold).  The norm and orthogonality follow  from the properties of the one particle spaces as above (with the rule that corresponding elements of the sequences are contracted by scalar product). We may therefore write
$$\psi_\tau (x_1, x_2,\dots x_N) = \Sigma a_{\alpha_1, \alpha_2 \dots \alpha_N}  \phi_{\alpha_1 , \tau}(x_1) \phi_{\alpha_2 , \tau}(x_2)\cdots \phi_{\alpha_N , \tau}(x_N). \eqno(8.1)$$
We now argue that a local unitary transformation of the form $e^{i\Lambda (x,\tau)}$ should act, with the {\it same} function $\Lambda(x,\tau)$ in each of the factor spaces.\footnote{*}{One can think of this procedure as the action of an {\it operator} $ \bigl(e^{i{\bf \Lambda}}\bigr)^N =\bigl(e^{i{\bf \Lambda}N}\bigr)$ acting on the $N$ particle state.} This construction provides a convenient mechanism for the gauge transformations of the Bose-Einstein or Fermi-Dirac Fock spaces (and, in general, for linear combinations). Furthermore, as we shall see below, it enables us to define a {\it field} $a_\mu(x,\tau)$.  We therefore define the gauge transformation $\psi \rightarrow \psi'$ as
$$\eqalign{\psi'_\tau (x_1, x_2,\dots x_N) &= \Sigma  a_{\alpha_1, \alpha_2 \dots \alpha_N}  e^{i(\Lambda(x_1 , \tau)+ \Lambda(x_2 , \tau) + \cdots \Lambda(x_N , \tau))}\cr
&\times \phi_{\alpha_1 , \tau}(x_1) \phi_{\alpha_2 , \tau}(x_2)\cdots, \phi_{\alpha_N , \tau}(x_N)\cr
&=e^{i(\Lambda(x_1, \tau)+ \Lambda(x_2, \tau) + \cdots \Lambda(x_N, \tau))} \psi_\tau (x_1, x_2,\dots, x_N)\cr} \eqno(8.2)$$
\par The remaining argument is the same as for the one body case. At each point $x_i$ of the wave function associated with the $i^{th}$ factor, we have
$$ \eqalign{\bigl(-i {\partial \over \partial {x_i}^\mu} - a'_\mu(x_i, \tau)\bigr) \psi'_\tau(x_1, x_2,\dots x_N) &= 
e^{i(\Lambda(x_1, \tau)+ \Lambda(x_2, \tau) + \cdots \Lambda(x_N, \tau))}\cr
&\bigl(-i {\partial \over \partial {x_i}^\mu} - a_\mu(x_i, \tau)\bigr) \psi_\tau(x_1, x_2,\dots x_N),\cr} \eqno(8.3)$$
so that
$$ a'_\mu(x_i,\tau) = a_\mu(x_i,\tau) + {\partial \Lambda(x_i,\tau) \over \partial {x_i}^\mu}, \eqno(8.4)$$
It therefore follows that this procedure leads to a local covariant {\it field} for the electromagnetic potential vector. 
\par If we call the compensation function for the $\tau$ evolution $a_5(x_1, x_2,\dots x_N, \tau)$ and
$$\Lambda(x_1, x_2,\dots x_N, \tau)= \Lambda(x_1, \tau)+ \Lambda(x_2, \tau) + \cdots \Lambda(x_N, \tau) \eqno(8.5)$$
then it follows that
$$ a'_5(x_1, x_2,\dots x_N, \tau)= a_5(x_1, x_2,\dots x_N, \tau)+ {\partial\over \partial \tau} \Lambda(x_1, x_2,\dots x_N,\tau). \eqno(8.6)$$

\par The fifth gauge function is clearly here not a property of the individual particles, and in this sense it appears not to correspond to a local field on the individual particles (it is a local field, however, on the configuration space $(x_1, x_2,\dots x_N)$).
\par The field strengths associated with the fifth field form a set
$${f^i}_{\mu 5} (x_1, x_2,\dots x_N) = \partial_{\mu i} a_5(x_1, x_2,\dots x_N) -\partial_\tau a_\mu(x_i). \eqno(8.7)$$
Under gauge transformation,
 $$ \eqalign{\partial_{\mu i}( a_5(x_1, x_2,\dots x_N, \tau)&+ {\partial\over \partial \tau} \Lambda(x_1, x_2,\dots x_N,\tau)) -\partial_\tau (a_\mu(x_i,\tau) + {\partial  \over \partial {x_i}^\mu} \Lambda(x_i \tau))\cr
 &=\partial_{\mu i} a_5(x_1, x_2,\dots x_N,\tau) -\partial_\tau a_\mu(x_i,\tau),\cr} \eqno(8.8)$$
 since the derivative ${\partial  \over \partial {x_i}^\mu}$ selects the term in the sum that cancels ${\partial  \over \partial {x_i}^\mu} \Lambda(x_i \tau)$.
 \par To be able to write the elements of this set in a uniform way in the arguments, we define a field on $x_1, x_2,\dots x_N$ for each $\tau$ such that the projection 
 $${a^i}_\mu (x_1, x_2,\dots x_N, \tau)= a_\mu(x_i,\tau). \eqno(8.9)$$
 We can then  write 
$${f^i}_{\mu 5} (x_1, x_2,\dots x_N) = \partial_{\mu i} a_5(x_1, x_2,\dots x_N) -\partial_\tau {a^i}_\mu(x_1, x_2,\dots x_N). \eqno(8.10)$$
\par We now discuss the electromagnetic current and field equations for the $N$ body case.
\par We write the action for an $N$ particle system in the presence of electromagnetism as
$$\eqalign{S &=\int d\tau \Pi_i d^4x_i \sqrt{g(x_i)} \bigl\{ i {\psi^*}_\tau(x_1,x_2,\dots x_N)\cr
&{\partial \over \partial \tau} \psi_\tau(x_1,x_2,\dots x_N) -i \psi_\tau(x_1,x_2,\dots x_N) {\partial \over \partial \tau} {\psi^*}_\tau(x_1,x_2,\dots x_N)\cr
&+ a_5(x_1,x_2,\dots x_N, \tau){\psi^*}_\tau(x_1,x_2,\dots x_N)\psi_\tau(x_1,x_2,\dots x_N) \cr
&-  {\psi^*}_\tau(x_1,x_2,\dots x_N)\Sigma_i \bigl\{(p_\mu-a_\mu(x_i,\tau)) g^{\mu \nu}(x_i)
( p_\nu-a_\nu(x_i,\tau))\cr & -a_5(x_1,x_2,\dots x_N,\tau)\bigr\}\psi_\tau(x_1,x_2,\dots x_N)\bigr\} \cr }\eqno(8.11)$$
\par As for the one-particle case, we add an electromagnetic part
$$ \eqalign{S_{em}&= +\int d\tau \Sigma_i \int d^4x_i  {1 \over 4\sqrt{g}}\bigl( f^{\mu \nu}(x_i,\tau) f_{\mu \nu}(x_i, \tau)\bigr\}\cr
&+{ f^i}_{\mu 5}(x_1,x_2,\dots x_N){f^i}^{\mu 5}(x_1,x_2,\dots x_N)\bigr).\cr} \eqno(8.12)$$ 
 The  variation with respect to $a_5$ (taking into account the factor $g^{\mu\nu}$ in raising the index) yields the equation of motion
$$ \eqalign{{f_{\mu 5}}^{;\mu}(x_1,x_2,\dots x_N,\tau)&= {\psi_\tau}^*(x_1,x_2,\dots x_N)\psi_\tau(x_1,x_2,\dots x_N)\cr
&= \Pi_i {g(x_i)}^{-{1\over 2}} \rho(x_1,x_2,\dots x_N,\tau)\cr} , \eqno(8.13)$$
the density on the full space $(x_1,x_2,\dots x_N)$ at each $\tau$. 
\par  Finally, the variation with respect to $a_\mu(x_i,\tau)$ of the interaction term, since
 $f_{\mu \nu} (x,\tau)$ is a one particle quantity, the field equations
 $$ {f_{\mu\nu}}^{;\nu}(x_i,\tau) = j_\mu (x_i,\tau), \eqno(8.14)$$
 where (the variation in $a_\mu(x_i,\tau)$ fixes $\tau$ but not the coordinates except for $x_i$)
 $$\eqalign{j_\mu (x_i,\tau)&= \Pi_{j\neq i} d^4x_j \sqrt{g(x_j)}\cr
 &{1 \over 2Mi\sqrt{g(x_i)}} \bigl({\psi_\tau}^*(x_1,x_2,\dots x_N) g^{\mu\nu}(x_i)(p_{\nu i} -ia_\nu(x_i))\psi_\tau(x_1,x_2,\dots x_N)\cr  &- ((p_{\nu i} +a_\nu(x_i)){\psi_\tau}^*(x_1,x_2,\dots x_N))g^{\mu\nu}(x_i)\psi_\tau(x_1,x_2,\dots x_N) \bigr).\cr} \eqno(8.15)$$ 
 \par The local one-particle field equations are then
 $$ {f_{\mu \nu}}^{; \mu} (x_i, \tau) =  j_\nu (x_i,\tau) \eqno(8.16)$$
 or, as in the one-particle case, 
$$\partial_{\mu i} (\sqrt{g(x_i)} f^{\mu \nu}(x_i,\tau) ) = \sqrt{g(x_i)}j^\nu(x_i, \tau).\eqno(8.17) $$
 \par The Lorentz force acting on a particle in this many body framework is then
$$ F^\mu (x_i, \tau) = {f^\mu}_\nu (x_i, \tau) {{dx_i}^\nu \over d\tau}, \eqno(8.18)$$
providing a basis, for example, for writing Vlasov equations in general relativistic statistical mechanics.
\bigskip
\noindent{\it 9. Summary and Outlook}
\bigskip
\par We have shown that the SHP theory can be embedded by local coordinate transformations into the framework of general relativity. The Minkowski spacetime coordinates of the SHP theory are considered to lie in the tangent space of a manifold with metric and connection form derived from the coordinate transformations on the equations of motion for particles moving on the locally flat Minkowski spacetime, parametrized by a universal monotonic world  time $\tau$. The four momentum is well-defined on the manifold, and a formula for its $\tau$ derivative, which may be understood as a ``force'', is obtained, displaying the effect of the potential as well as the curvature (through the connection form). The canonical momentum vector $p_\mu$ in the cotangent space has simple canonical Poisson brackets with the coordinates $x^\mu$, but we show that it is the mapping $p^\mu = g^{\mu\nu}(x) p_\nu$ back to the tangent space which corresponds to the measured energy and momentum, and compute the energy and momentum of a particle near the Schwarzschild radius (horizon) of a black hole.
\par For the many body system, each particle, at the points $\{ {x_i}^\mu\}$,  is assumed to move locally on a flat Minkowski space, which is then transformed by local coordinate transformation to the manifold of GR with coordinates ${x_i}^\mu$. Since particles with (flat Minkowski) coordinates $\xi_1,\xi_2 \dots \xi_N$ lie in different local tangent spaces at the points      
$x_1,x_2 \dots x_N$ of the curvilinear coordinatization of GR, Poincar\'e invariance of the potential function is not applicable.
\par Since the Poisson bracket of the SHP theory is unchanged in form under local diffeomorphisms, it forms the basis of a quantum theory in which the momentum operator generates infinitesimal translations along the local coordinates. However, the operator $-i {\partial\over \partial x^\mu}$  is not Hermitian on a Hilbert space of functions $\psi_\tau(x)$, square integrable over the invariant measure $d^4x \sqrt{g(x)}$. It was necessary to define the Hermitian momentum operator $ p_\mu = -i {\partial \over \partial x^\mu} - {i \over 2}{\partial \over \partial x^\mu} \sqrt {g(x)}$, in analogy to the operator defined by Newton and Wigner [52] in momentum space. We showed, in the discussion of Fourier transforms, that this operator generates infinitesimal translations. 
\par We then developed the basic scattering theory in this quantum mechanical framework. The interaction picture expansion for a potential model is worked out, similar to the Feynman type expansions, but with more complicated vertices due to the curvature of space time. We also showed that the propagator for ``free'' evolution (Green's function) has a Laplace-Beltrami operator in the denominator which, for small curvature, reduces to the flat space SHP Hamiltonian with the addition of an effective mass shift due to the curvature.    
\par This Hilbert space provides a basis for a local $U(1)$ gauge, for which the compensation fields (sections on the bundle) correspond to classical $5D$ electromagnetic fields [59][60]. We obtain field equations for the electromagnetic fields and associated currents from an action ($\tau$ integrated Lagrangian).
\par The many body quantum theory is treated by constructing a tensor product space and the associated electromagnetic theory is developed assuming that each factor in the tensor product carries the same gauge transformations.  This enables us to define a gauge compensation field $a_\mu (x)$ for the four components which can be evaluated on each particle, but due to the universality of $\tau$, the fifth component must depend on the coordinates of all of the particles as a locally defined function on the full configuration space, similar to the function $V(x_1,x_2,\dots x_N)$ of the potenial model.
\par The work of this paper is primarily restricted to describing a relativistic dynamics in a $\tau$ independent gravitational field, {\it i.e.}, the metric is assumed independent of $\tau$.  Since the connection has the same structure as in GR, one can write Einstein's equations in the same form ({\it e.g.}[64]). Therefore, in this case,the energy momentum tensor, determining $g^{\mu\nu}(x)$, should be independent of $\tau$. To achieve this, one may use partially integrated currents, taking into account correlations [65], or the zero modes extracted from full integration yielding $4D$ conserved currents [1][59][60]. In the more general case, where the structure of spacetime is dynamical (for example, star formation, collision between stars or black holes, or for unstable stars such as supernova) the energy momentum tensor would depend on $\tau$. We show in the Appendix how, in such cases, the corresponding explicit dependence of the local transformations from the Minkowski space coordinates to the curved coordinates can be expressed in terms of such a $\tau$ dependent metric tensor.  
\par The classical results of this paper provide an eight dimensional phase space for general relativity, just as the SHP theory provides for special relativity, and therefore a general relativistic statistical mechanics can be formulated. The assumptions necessary to construct Gibbs ensembles [66][67] in this context must be carefully examined; we leave this subject for a future publication.
\par  The  many body problem, generalizing the work of Horwitz and Arshansky [68], can be formulated in this framework. The many body Hilbert space can be used to construct a Fock space as the basis for a quantum field theory.
\par We finally remark that the results of the work of this paper provide a basis for the vector field approach of Bekenstein and Sanders[21] and the associated discussion of nonabelian gauge fields given by Horwitz, Gershon and Schiffer [19]; it will therefore be of interest to follow the development of the $U(1)$ gauge theory given here with a study of non-Abelian gauge theories (see [60]). 
\bigskip
\bigskip
\bigskip
\noindent{\it Appendix}
\bigskip
\par We study here the effect of a $\tau$ evolving spacetime, a situation which would occur if the energy momentum tensor depends on $\tau$. 
\par Generally, at a point $x^\lambda$, the velocity of a particle is ${\dot x}^\lambda$, just a motion on the coordinates $\{ x\}$. If the spacetime is changing, we think of the tangent space as reflecting this change. Therefore, the local coordinatization $\xi$ changes as the world coordinates evolve. At each $\tau$, it is still true that
$$ d\xi^\mu = {\partial \xi^\mu \over \partial x^\lambda} dx^\lambda, \eqno(A.1)$$
but ${\partial \xi^\mu \over \partial x^\lambda}$ changes as the particle moves and as the spacetime evolves. We can write
$$ {\partial \xi^\mu \over \partial x^\lambda} = {\partial \xi^\mu \over \partial x^\lambda}(x(\tau), \tau) \eqno(A.2)$$
so that
$$ {d \over d\tau}{\partial \xi^\mu \over \partial x^\lambda} = {\partial^2 \xi^\mu \over \partial x^\lambda \partial x^\sigma }((x(\tau),\tau) {\dot x}^\sigma + {\partial \over \partial \tau}{\partial \xi^\mu \over \partial x^\lambda} (x(\tau), \tau), \eqno(A.3)$$
where the second term is due to the change in orientation of the $\xi^\mu$ coordinates in $\tau$.
\par The canonical structure postulated in $(1.1)$ and $(1.2)$, and the definition of $g_{\mu \nu}$ remain the same but, as a function of  ${\partial \xi^\mu \over \partial x^\lambda}(x,\tau)$, it now becomes a function of $\tau$.  We therefore have
$$K =  {M\over 2} g_{\mu\nu}(x(\tau), \tau){\dot x}^\mu {\dot x}^\nu + V(x). \eqno(A.4)$$
We now calculate from $(A.1)$ the total $\tau$ derivative
$$ \eqalign{{\ddot \xi}^\mu &= {d \over d\tau} \bigl(  {\partial \xi^\mu \over \partial x^\lambda}{\dot x}^\lambda \bigr)\cr
&= {\partial^2 \xi^\mu \over \partial x^\lambda \partial x^\gamma} {\dot x}^\gamma {\dot x}^\lambda \cr
&+ {\partial \over \partial \tau}{\partial \xi^\mu \over \partial x^\lambda} {\dot x}^\lambda + {\partial \xi^\mu \over \partial x^\lambda} {\ddot x}^\lambda  \cr
&= - { 1\over M} \eta^{\mu\nu}{\partial x^\lambda\over \partial \xi^\nu}{\partial V(x)\over \partial x^\lambda}, \cr} \eqno(A.5)$$
where the last equality, as before, follows from the canonical structure of the tangent space. Muliplying by ${\partial x^\sigma \over \partial \xi^\mu}$, and solving for ${\ddot x}^\sigma$, we find a geodesic type equation as before but with an additional (velocity dependent) term  
$$  {\ddot x}^\sigma = - {\Gamma^\sigma}_{\lambda \gamma} {\dot x}^\gamma {\dot x}^\lambda
  - { 1\over M} g^{\sigma \lambda} {\partial V(x)\over\partial x^\lambda} - {\partial x^\sigma \over \partial \xi^\mu} {\partial \over \partial \tau}\bigl({\partial \xi^\mu \over \partial x^\lambda}\bigr) {\dot x}^\lambda.  \eqno(A.6))$$
  \par We now show that $ {\partial \over \partial \tau}\bigl({\partial \xi^\mu \over \partial x^\lambda}\bigr)$ can be expressed in terms of ${\partial \over \partial \tau} g_{\mu\nu} (x(\tau),\tau)$, which then carries the information, from the Einstein equations, about the evolution of the spacetime.
  \par From the definition $(1.9)$ we compute
  $$\eqalign{{\partial g_{\mu \nu}  \over \partial \tau}(x(\tau),\tau) &= \eta_{\sigma \gamma} \bigl[ {\partial \over \partial \tau} \bigl({\partial \xi^\sigma \over \partial x^\mu}\bigr) {\partial \xi^\gamma \over \partial x^\nu}\cr &+ {\partial \xi^\gamma \over \partial x^\mu}{\partial \over \partial \tau} \bigl({\partial \xi^\sigma \over \partial x^\nu}\bigr) \bigr], \cr}\eqno(A.7)$$
where we have used the $\sigma,\gamma$ symmetry of $\eta_{\sigma \gamma}$ in the second term.
\par Now, define
$${\partial \over \partial \tau} \bigl({\partial \xi^\mu \over \partial x^\lambda}\bigr) \equiv {t^\mu}_\lambda . \eqno(A.8)$$
Then we can write $(A.7)$ as
  $$\eqalign{{\partial g_{\mu \nu}  \over \partial \tau}(x(\tau),\tau) &= \eta_{\sigma \gamma} \bigl[ {t^\sigma}_\mu {\partial \xi^\gamma \over \partial x^\nu} + {t^\sigma}_\nu {\partial \xi^\gamma \over \partial x^\mu} \bigr] \cr
  &=\eta_{\sigma \gamma} \bigl({\delta_\mu}^\lambda {\partial \xi^\gamma \over \partial x^\nu} + {\delta_\nu}^\lambda {\partial \xi^\gamma \over \partial x^\mu} \bigr) {t^\sigma}_\lambda \cr
  &\equiv {{M_{\mu \nu}}^\lambda}_\sigma  {t^\sigma}_\lambda. \cr} \eqno(A.9)$$
  This equation can be inverted with the matrix ${N_\lambda}^{\sigma \mu \nu} $ satisfying
  $$  {N_{\lambda '}}^{\sigma' \mu \nu} {{M_{\mu \nu}}^\lambda}_\sigma = {\delta_{\lambda'}}^\lambda {\delta_\sigma}^{\sigma'}. \eqno(A.10)$$
It is a possible exceptional case that ${N_\lambda}^{\sigma \mu \nu} $ could be singular; this would correspond to a singular development of the transformation function ${\partial \xi^\sigma \over \partial x^\nu}$ in $\tau$, which we do not treat here.
\par Multiplying $(A.9)$ by ${N_{\lambda '}}^{\sigma' \mu \nu}$, we obtain
$$ {N_{\lambda '}}^{\sigma' \mu \nu} {\partial g_{\mu \nu}  \over \partial \tau}(x(\tau),\tau) = {t^{\sigma'}}_{\lambda'} = {\partial \over \partial \tau} \bigl({\partial \xi^{\sigma'} \over \partial x^{\lambda'}}\bigr). \eqno(A.11)$$
The solution that we have found, as well as many of the relations we have used in our study, involve the transformation functions ${\partial \xi^\gamma \over \partial x^\mu}$; as pointed out by 
Weinberg [64], these functions enter quadratically in $g_{\mu \nu}$, and are therefore determined
up to Lorentz transformation, at any given $\tau$, by $g_{\mu \nu}$.
\bigskip

\noindent{\it Acknowledgements}
\par I wish to thank Asher Yahalom, Yossi Strauss, Igal Aharonovich, Gil Elgressy and Martin Land for helpful discussions. 

\bigskip
\noindent{\it References}
\frenchspacing
\bigskip

\item{1.} Lawrence Horwitz, {\it Relativistic Quantum  Mechanics},  Fundamental Theories of Physics 180, Springer, Dordrecht (2015).
\item {2.} E.C.G. Stueckelberg, Helv. Phys. Acta {\bf 14}, 372 (1941).
\item{3.} E.C.G. Stueckelberg, Helv. Phys. Acta {\bf 14},585 (1941).
\item{4.} E.C.G. Stueckelberg, Helv. Phys. Acta {\bf 15}, 23 (1942).
\item{5.} L.P. Horwitz and C. Piron, Helv. Phys. Acta {\bf 66}, 316
(1973).
\item{6.} R.E. Collins and J.R. Fanchi, Nuovo Cim. {\bf 48A},
314 (1978).
\item{7.} J.R. Fanchi, {\it Parametrized Relativistic Quantum
Theory}, Kluwer, Dordrecht (1993).
\item{8.}  Isaac Newton, {\it Philosophia Naturalis
Principia Mathematica},
London 1687.
\item{9.} I.B. Cohen and A. Whitman  {\it The
Principia: Mathematical Principles of Natural Philosophy: A  New
Translation}, University of California Press, Berkeley (1999).
\item{10.} R.I. Arshansky and L.P. Horwitz, Jour. Math. Phys. {\bf 30},
66 (1989).
\item{11.}  R.I. Arshansky and L.P. Horwitz, Jour. Math. Phys. {\bf 30} 380 (1989).
\item{12.} R.I. Arshansky and L.P.Horwitz, Jour. Math. Phys. {\bf 30}, 213 (1989).
\item{13.} J. Schwinger, Phys. Rev. {\bf 73}, 416 (1948).
\item{14.}  J. Schwinger,Phys. Rev. {\bf 74}, 1439 (1948).
\item{15.} S. Tomonaga, Phys. Rev {\bf 74}, 224 (1948).
\item{16.} N.E.J. Bjerrum-Bohr, P.H. Damgaard, G. Festuccia, L. Plante and P. Vanhove,  arXiv:1806.04920 [hep-th](2018).
\item{17.}T. Damour, Phys. Rwev. D {\bf 98}, 104015 (2018).
\item{18.}T. Damour,  Phys. Rev. D {\bf 97,} 044038 (2018).
\item{19.} L.P. Horwitz, A. Gershon and M. Schiffer, Found. Phys. {\bf 41},
141 (2010).
\item{20.} A. Gershon and L.P. Horwitz, Jour. Math. Phys. {\bf 50}, 102704 (2009).
\item{21.}J.D. Bekenstein and R.H. Sanders, Astrophys. Jour. {\bf 429}, 480 (1994).
\item{22.}  J.D. Bekenstein and M. Milgrom, Astrophys. Jour.{\bf 286}, 7 (1984). 
\item{23.} J.D. Bekenstein, Phys. Rev. {\bf D70}, 083509 (2004).
\item{24.} J.D. Bekenstein, Contemporary Physics {\bf 47} 387 (2006) for a review, references, and further development.
\item{25.} M. Milgrom, Asrophys. Jour. {\bf 270},365 (1983).
\item{26.} M. Milgrom, Asrophys. Jour. {\bf 270},371(1983)
\item{27.} M. Milgrom, Asrophys. Jour. {\bf 270},384 (1983).
\item{28.} N.B. Birrell and P.C.W. Davies, {\it Quantum Fields in Curved Space}, Cambrdge Monographs on Mathematical Physics, Cambridge (1982).
 \item{29.} E. Poisson, {\it A Relativist's Toolkit}, Cambridge (2004).
 \item{30.} Ana Cannas de Silva, {\it Lectures on Symplectic Geometry}, Lecture Notes in Mathematics 1764, Springer (2006).
 \item{31.} P.A.M. Dirac, {\it Quantum Mechanics}, 1st edition ,Oxford University Press, London (1930);3rd edition (1947).
 \item{32.} L. Van Hove,  Proc. Roy. Aca. Belgium {\bf 26}, 1 (1951).
 \item{33.} H.J. Groenwold, Physica {\bf 12} 405 (1946).
\item{34.} L.P. Horwitz, Y. Ben Zion, M. Lewkowicz, M. Schiffer and J.Levitan, Phys. Rev. Lett. {\bf 98} 234301 (2007).
\item{35.} M.D. Kruskal, Phys. Rev. {\bf 119} 1743.
\item{36.} K. Schwarzschild, Sitzungsberichte der K\"oniglich Preussischen Akademie der Einsteinschen Theorie {\bf 1}:424 (1916).
\item{37.} P.A.M. Dirac, {\it General Theory of Relativity}, Wiley, New York (1975).
\item{38.} S. Hawking, Nature {\bf 248} 30 (1974). 
\item{39.} D. Ludwin and L.P. Horwitz, Jour. Math. Phys.{\bf 52} 012303 (2011).
\item{40.} D. Momemi, {\it Exact classical and quantum solutions for a covariant oscillator near the black hole horizon in he Stueckelbederg-Horwitz-Piron theory},  to be published in Phys. Lett. A.
\item{41.} A. Friedman. Zeits. fur Physik {\bf A 21}, 326
(1924)
\item{42.} G. Lemaitre, Monthly Notices of the Roy. Astro. Soc. {\bf 91},
483 (1931)[trans. from Ann. de la Societe Scientifique de Bruxelles
{\bf A 47} 49 (1927)].
\item{43.} G. Lemaitre, Ann. de la Societe
Scientifique de Bruxelles {\bf A 53}, 51 (1933).
\item{44.} H.P. Robertson,
Astrophys. Jour. {\bf 82}, 284 (1935).
\item{45.} H.P. Robertson,Astrophys. Jour. {\bf 83}, 187 (1936).
\item{46.} H.P. Robertson,Astrophys. Jour. {\bf 83}, 257 (1936).
\item{47.} A.G. Walker, Proc. London Math. Soc.  {\bf 42} 90 (1936).
\item{48.} J. Schwinger, Phys. Rev. {\bf 82}, 664 (1951).
\item{49.} B.S. DeWitt, Physics Reports {\bf 19}, 295 (1975).
\item{50.} B.S. DeWitt, {\it The Global Approach to Quantum Field Theory}, Oxford University Press, Oxford (2002).
\item{51.} S.S. Schweber, {\it An Introduction to Quantum Field Theory}, Harper and Row, New York (1964), pages 419, 420.   
\item{52.} T.D. Newton and E.Wigner, Rev. Mod. Phys. {\bf 21}, 400 (1949).
\item{53.} L.L. Foldy and S.A. Wouthuysen. Phys. Rev. {\bf 78}, 29 (1950).
\item{54.} R.P. Feynman, Phys. Rev. {\bf 76},  749 (1949).
\item{55.} R.P. Feynman, Phys. Rev. {\bf 76},  769 (1949).
\item{56.} J. Schwinger, Proc. Nat. Aca. Science {\bf 37}, 452 (1951).
 \item{57.} S. Helgason, Groups and Geometric Analysis, Academic Press, New York (1984).
\item{58.} C.N. Yang, Annals of the New York Academy of Science {\bf 294} 86 (1977).
\item{59.}D. Saad, L.P. Horwitz and R.I. Arshansky, Foundations of Physics, {\bf 19} 1125 (1989).  
\item{60.} M.C. Land, N. Shnerb and L.P. Horwitz, Jour. Math. Phys. {\bf 36}, 3263 (1995).
\item{61.} J.D. Jackson, {\it Classical Electrodynamics},
2nd edition, John Wiley and Sons, New York (1974). 
\item{62.} S.A. Hojman and L.C. Sheply, Jour. Math. Phys. {\bf 32}, 142 (1991).
\item{63.} S. Tanimura, Ann. Phys. {\bf 220}229 (1992).
\item{64.} S. Weinberg, {\it Gravitation and Cosmology}, John Wiley and Sons, New York (1972).
\item{65.} M. Land, IOP Conference Series:Jour. of Physics Conf. Series {\bf 845}, 012024 (2017).
\item{66.}  L.P. Horwitz, W.C. Schieve and C. Piron, Ann. of
Phys. {\bf 137}, 306 (1981).
\item{67.} L.P. Horwitz, S. Shashoua and
W.C. Schieve, Physica {\bf A 161}, 300 (1989).
\item{68.} Lawrence P. Horwitz and Rafael I. Arshansky, {\it Relativistic Many-Body Theory and Statistical Mechanics}, IOP Concise Physics, A. Morgan and Claypool, Bristol (2018).

\end